\documentclass[runningheads]{llncs}
\usepackage[T1]{fontenc}
\usepackage{color}
\usepackage{amsmath}
\usepackage{booktabs}
\usepackage{tabularx}
\usepackage{siunitx}
\usepackage{soul}

\usepackage{csquotes}
\usepackage{hyperref}

\usepackage{graphicx}
\usepackage{fancyhdr}
\usepackage{etoolbox}
\AtBeginEnvironment{tabular}{\scriptsize}
\AtBeginEnvironment{tabularx}{\scriptsize}

\usepackage{array}
\newcolumntype{Y}{>{\raggedright\arraybackslash}X}

\begin{document}
\title{Increase Alpha: Performance and Risk of an AI-Driven Trading Framework}
\titlerunning{AI-Driven Trading Framework}
%
\author{Sid Ghatak\inst{1} \and Arman Khaledian\inst{2} \and Navid Parvini\inst{2} \and Nariman Khaledian\inst{2}}

\authorrunning{S. Ghatak et al.}

\institute{
Increase Alpha, LLC\\
\email{s.ghatak@increasealpha.com}
\vspace{0.2cm}
\and
Zanista AI Ltd.\\
\email{arman.khaledian@zanista.ai, nariman.khaledian@zanista.ai, navid.parvini@zanista.ai}
}

\maketitle              

\begin{abstract}

There are inefficiencies in the financial market, leaving unexploited patterns embedded in price, volume, and cross-sectional relationships. While recent advances increasingly employ large-scale transformer architectures, we take a domain-focused route: classical feed-forward and recurrent networks paired with expertly curated features to mine subtle regularities in noisy financial data. This smaller-footprint design is computationally lean and reliable under low signal-to-noise conditions—qualities that matter for daily production at scale. At \emph{Increase Alpha}, we developed a deep-learning framework that maps a universe of over 800 U.S. equities into daily directional signals with minimal computational overhead. 

The purpose of this paper is twofold. First, we outline the general overview of the predictive model without disclosing its core underlying concepts. Second, we evaluate its real-time performance through transparent, industry standard metrics. Forecast accuracy is benchmarked against both naive baselines and macro indicators. The performance outcomes are summarized via cumulative returns, annualized Sharpe ratio, and maximum drawdown. The best portfolio combination using our signals provides a low-risk, continuous stream of returns with a Sharpe ratio of more than 2.5, maximum drawdown of around 3\%, and a near-zero correlation with the S\&P 500 market benchmark. We also compare the model's performance through different market regimes, such as the recent volatile movements of the US equity market in the beginning of 2025. Our analysis showcases the robustness of the model and significantly stable performance during these volatile periods.

Collectively, these findings show that market inefficiencies can be systematically harvested with modest computational overhead if the right variables are considered. This report will emphasize the potential of traditional deep learning frameworks for generating an AI-driven edge in the financial market.

\end{abstract}

\section{Introduction}
Financial markets are often portrayed as an example of informational efficiency, a view that traces its intellectual lineage to the early‐twentieth-century exchange between Louis Bachelier and Henri Poincaré. While reviewing Bachelier’s 1900 doctoral thesis, Poincaré remarked that if security prices truly followed an unrestricted random walk, the expectation of profit “could not rationally differ from zero.” Many decades later, Eugene Fama stated that intuition into the Efficient Market Hypothesis (EMH), arguing that all available information is instantaneously and correctly realized into prices. Yet the empirical literature now documents persistent pockets of predictability based on momentum, reversal, seasonal effects, and cross-sectional valuation that careful observers can still exploit.

In recent years, artificial intelligence has emerged as a transformative force in quantitative finance, lowering barriers to entry and enabling independent researchers and small trading teams to harness advanced methodologies once exclusive to major institutions. By leveraging open‐source libraries and cloud computing platforms, practitioners can now prototype, validate, and deploy AI‐powered strategies across diverse market regimes with unprecedented ease. This democratization of technology has given rise to a rich ecosystem of algorithms that span reinforcement learning, deep learning, ensemble models, and evolutionary techniques, each offering unique avenues for extracting predictive signals from complex financial data. For example, high‐frequency trading now represents over 70 \% of all executed trades \cite{cohen2022algorithmic}, and modern HFT engines rely critically on machine‐learning and deep‐learning methods to scan multiple data streams, generate sub‐second buy/sell signals, and execute vast numbers of orders \cite{huang2019automated,mcgroarty2019high}. Some implementations focus on drawdown control—e.g.\ a grid‐trading DL system that adaptively reduces downside risk \cite{rundo2019grid}—while others apply neural architectures to volatile assets such as Bitcoin, leveraging parallel processing and Bayesian regularization to forecast returns in noisy, high‐frequency environments \cite{lahmiri2020intelligent}.

Reinforcement learning agents frame trading as a sequential decision‐making problem, iteratively learning to select buy, hold, or sell actions that maximize cumulative reward under evolving market conditions. While these methods can uncover highly nonlinear policies that adapt to regime shifts, they require vast historical datasets and significant computational resources and carry a pronounced risk of overfitting unless subjected to rigorous backtesting and paper‐trading validation \cite{bai2025review}. Complementing this, Long Short‐Term Memory (LSTM) networks excel at modeling temporal dependencies in price series, capturing both seasonality and momentum trends; however, they remain vulnerable to unanticipated “black swan” events and mandate careful feature engineering—integrating technical indicators alongside raw OHLCV data—and walk‐forward validation to ensure robustness \cite{song2024lstm}.

Supervised timing models—most notably tree‐based ensembles such as Random Forests and XGBoost—classify market regimes by leveraging a curated set of features that may include momentum scores, volatility measures, and sentiment indices. Their interpretability via feature‐importance metrics facilitates systematic refinement, yet they also demand prudent cross‐validation protocols to avoid capitalizing on spurious historical patterns \cite{simsek2024improving}. In parallel, Generative Adversarial Networks (GANs) offer a novel data‐augmentation paradigm, synthesizing realistic time‐series scenarios to fill gaps in rare or stressed‐market conditions; despite their promise in mitigating overfitting, GANs introduce training instability and necessitate thorough statistical validation of the synthetic output against empirical distributions \cite{vuletic2024fin}.

Natural‐language processing pipelines extend the AI toolkit by quantifying sentiment from unstructured text sources—news articles, social‐media posts, and corporate filings—thus capturing behavioral drivers that often precede price movements. Beginning with lexicon‐based scoring and progressing to fine‐tuned transformer architectures, these systems must contend with evolving language trends, sarcasm, and manipulation, calling for continuous retraining and robust noise‐filtering mechanisms \cite{rahimikia2024r}. Similarly, multi‐factor AI models synthesize fundamental, technical, and alternative datasets to rank and select assets; machine‐learning algorithms optimize factor combinations and, through systematic rebalancing and transaction‐cost modeling, seek to sustain persistent outperformance while managing factor decay.

Dynamic portfolio allocators harness Bayesian inference or reinforcement learning to adaptively balance risk and return, continuously updating asset weights in response to shifting volatility and correlation structures. By incorporating frictional cost simulations and liquidity constraints into backtests, these systems aim to enhance risk‐adjusted performance while controlling turnover. Across all methodologies, the pathway to robust deployment consistently emphasizes thorough backtesting, staged paper trading, and ongoing out‐of‐sample monitoring, thereby laying the groundwork for sustainable, data‐driven trading paradigms in the era of AI. AI‐driven trading frameworks typically fuse three pillars of information:
\begin{itemize}
  \item \textbf{Technical analysis}: historical price, volume, and order‐book dynamics  
  \item \textbf{Fundamental analysis}: the impact of corporate and macroeconomic news on asset valuations  
  \item \textbf{Investor sentiment}: emotion and opinion signals mined from social media and news feeds  
\end{itemize}

At \emph{Increase Alpha}, we believe that market inefficiencies can be harvested systematically by using the information mentioned above mixed with classical feed‐forward and recurrent neural networks curated by domain experts. The use of transformers and large language models in Finance has gained popularity among the experts in the field in recent years \cite{lopez2023can,sarkar2024lookahead,rahimikia2024r}. However, in \emph{Increase Alpha}, rather than pursuing massive transformer models on unstructured text, our minimalist design extracts only the variables that a seasoned fundamental analyst would consider economically meaningful, and then uses compact networks to uncover non-linear interactions at scale. The result is a daily, security‐level directional signal for 814 U.S. equities—generated with negligible computational overhead and latency compatible with both discretionary and algorithmic execution.

Because the model and feature definitions are proprietary, we treat the architecture as a black box. In Section \ref{sec:methodology} we describe our data collection, signal generation, and execution infrastructure. In Section \ref{sec:accuracy} we benchmark performance—using cumulative return, annualized Sharpe ratio, and maximum drawdown—against naive baselines and macroeconomic indicators, and examine behavior across different market regimes, including the turbulent U.S. equity markets of early 2025.

The remainder of the paper is organized as follows:

\begin{itemize}
    \item Chapter 2 — Methodology
    \begin{itemize}
        \item Describes the signal-generation pipeline, including multi-horizon forecasting, timestamp-tracked storage, and implementation mechanism.
        \item Explains how execution logic is calibrated using hyperparameter grid search for Profit-Taker, Stop-Loss, and Maximum Holding Period.
        \item Outlines the cloud-based deployment on Azure AKS and the associated scalability and cost-efficiency considerations.
    \end{itemize}
    
    \item Chapter 3 — Accuracy and Statistical Significance
    \begin{itemize}
        \item Measures signal performance across 814 tickers using multiple holding horizons and directional classifications (long, short).
        \item Applies statistical rigor, including z-tests and confidence intervals, to validate signal effectiveness against random baselines.
        \item Summarizes signal reliability by evaluating coverage, profitability, and sample-size-adjusted significance.
    \end{itemize}

    \item Chapter 4 — Risk and Return
    \begin{itemize}
        \item Translates signal outputs into cumulative return/PnL, drawdown, and Sharpe ratio metrics using the optimized execution configuration.
        \item Compares the strategy’s performance against a buy-and-hold baseline and macro indices such as the S\&P 500.
        \item Highlights robustness under stress through a regime-based analysis—contrasting pre- and post-January 2025 performance—and visualizes results.
    \end{itemize}

    \item Chapter 5 — Portfolio Construction and Dynamic Rebalancing
    \begin{itemize}
        \item Converts signals into a real-world trading strategy using dynamic stock selection and quarterly rebalancing.
        \item Portfolio configurations and evaluates performance via P\&L, Sharpe ratio, and drawdown.
        \item Showcasing the adaptability of signals across regimes, confirming the real-world viability of the approach.
    \end{itemize}
    
    \item Chapter 6 — Conclusion
    \begin{itemize}
        \item Summarizes the predictive power, statistical rigor, and market resilience of the \emph{Increase Alpha} system.
        \item Emphasizes that classical deep learning when paired with expert feature design and scalable infrastructure, offers a viable and interpretable alternative to more opaque architectures in financial signal generation.
    \end{itemize}
\end{itemize}

\section{Methodology}\label{sec:methodology}
Our objective in this chapter is to discuss—without revealing proprietary intellectual property—the full life-cycle by which the \emph{Increase Alpha} framework generates, stores, filters, and evaluates daily equity--level trading signals. We also describe the metrics we used to evaluate the performance of the generated signals.  Every step is designed to eliminate \emph{look-ahead bias} and \emph{information leakage}, which are the main pitfalls in any trading strategies.  This section is organized as follows:

\begin{enumerate}
    \item Data-generation and signal specification (directional forecasts, storage, and tradable conventions)  
    \item Scenario analysis and cloud infrastructure (trading parameters search for profit-taker, stop-loss, and maximum holding period)
    \item Evaluation framework (accuracy tests, economic metrics, and robustness checks)  
\end{enumerate}

\subsection{Data-generation and signal specification}

Since \textbf{28 June 2021}, the production system has executed a daily inference cycle for a universe of 814 U.S. equities. Each trading day, shortly after the market close, the latest data—including official corporate actions, fundamental metrics, and price/volume features—are ingested and processed by our deep learning model.

The model generates ten distinct directional predictions per stock for the upcoming ten trading sessions (i.e., $t+1$ to $t+10$). Each signal forecasts the expected percentage price change for a specific future session, forming a sequence of forecasts from a single prediction date. For example, the prediction on 28 June 2021 includes 10 separate directional estimates for each of the trading sessions from 29 June to 13 July 2021.

Each forecast is stored along with two immutable timestamps, capturing the creation and final update moments of each signal. Every prediction is finalized only after the market close, using solely available data, and is time-stamped and committed before the opening of the target session. This rigorous tracking ensures a true ex-ante prediction record, immune to post-hoc adjustment or backtest contamination, and serves as a strong guardrail against \emph{look-ahead bias} and \emph{information leakage}. Each model run generates a rolling set of multi-horizon signals—ten predictions that span the next ten trading sessions—forming a comprehensive forward-looking view. This automated pipeline has been running uninterrupted since June 2021, producing a longitudinal dataset with consistent daily operations.

We source price data from \href{www.eodhd.com}{www.eodhd.com}, a comprehensive market data provider. Their paid subscription gives access to a wide range of live and historical market information. We collected daily OHLC prices for all 814 U.S. equities from 28 June 2021 to 30 June 2025.

The automated system has incurred a consistent operational cost since inception. Between November 2024 and April 2025, daily AWS compute expenses averaged approximately \$95--\$100 USD, with occasional spikes on high-load processing days. When including all essential infrastructure services—such as EC2 instances, storage (S3), RDS databases, SageMaker for model hosting, and supporting utilities—the total infrastructure expenditure for the period amounted to approximately \$17,000 USD. Notably, this excludes optional analytics services (e.g., QuickSight, LIT for Traders), which were not part of the inference or training pipeline. This steady and predictable cost profile underscores the system’s viability for long-term deployment and financial forecasting operations.

Table \ref{tbl:multi_horizon} presents an example prediction table for AAPL on 28 June 2021:

\begin{table}[htbp]
\centering
\caption{Example multi-horizon-ahead predictions with timestamp tracking}
\label{tbl:multi_horizon}
\begin{tabularx}{\textwidth}{p{3cm}YYYY}
\toprule
\textbf{Current Date-time} & \textbf{Ticker} & \textbf{Target Date} & \textbf{Forecasted return} & \textbf{Horizon} \\
\midrule
2021/06/28 - 21:30    & AAPL            & 29/06/2021           & +0.5835                   & 1 \\
2021/06/28 - 21:30    & AAPL            & 30/06/2021           & -3.5856                  & 2 \\
2021/06/28 - 21:30    & AAPL            & 01/07/2021           & +1.1635                   & 3 \\
2021/06/28 - 21:30    & AAPL            & 02/07/2021           & -1.2820                  & 4 \\
2021/06/28 - 21:30    & AAPL            & 06/07/2021           & -0.5109                  & 5 \\
2021/06/28 - 21:30    & AAPL            & 07/07/2021           & -0.5405                  & 6 \\
2021/06/28 - 21:30    & AAPL            & 08/07/2021           & -0.2841                  & 7 \\
2021/06/28 - 21:30    & AAPL            & 09/07/2021           & -0.3977                  & 8 \\
2021/06/28 - 21:30    & AAPL            & 12/07/2021           & -0.4024                  & 9 \\
2021/06/28 - 21:30    & AAPL            & 13/07/2021           & -0.2335                  & 10 \\
\bottomrule
\end{tabularx}
\end{table}

For each ticker the engine outputs a predicted return. We use the sign of these predicted returns as \emph{ternary direction code}
$s\in\{+1,0,-1\}$:

\begin{itemize}
    \item $+1$ (Long)  — expected positive open-to-close return next session,
    \item $-1$ (Short) — expected negative return,
    \item $0$ (Flat)   — neutral or low-confidence regime.
\end{itemize}

\subsection{Scenario Analysis and Trading Parameter Optimisation} \label{sec:scenario_analysis_and_optimization}

To convert directional signals into economic value, we extract three parameters necessary for trade execution.

\begin{itemize}
    \item \textbf{Profit-Taker (PT)} — positive return threshold for exit;
    \item \textbf{Stop-Loss (SL)} — adverse move forcing liquidation;
    \item \textbf{Maximum-Holding Period (MHP)} — maximum duration (days) a position may remain open.
\end{itemize}

To obtain these values, we design a backtesting algorithm that simulates real-world trading on historical data. We then conduct a grid search as a scenario analysis. For each ticker, we store the performance of each item in the signal universe under different PT, SL, and MHP settings. For each ticker, we run the scenario analysis over:

\begin{itemize}
    \item \textbf{Maximum Holding Period (MHP)}: ranges from 1 to 10;
    \item \textbf{Profit-Taker (PT)}: ranges from 0.001 to 0.02 in increments of 0.0005;
    \item \textbf{Stop-Loss (SL)}: ranges from $-0.04$ to $-0.01$ in increments of 0.005.
\end{itemize}

In each scenario, we use the signal directions and the prices below to compute the trade return $r$, Sharpe ratio (SR), and maximum drawdown (MDD). For day $t$ and ticker $i$, let
\[
    \operatorname{r}^{\text{}}_{i,t} =
    \begin{cases}
        \text{PT}, & \text{if } H_{i,t:t+MHP} - O_{i,t} \ge \text{PT},\\[4pt]
        \text{SL}, & \text{if } L_{i,t:t+MHP} - O_{i,t} \le \text{SL},\\[4pt]
        C_{i,t+MHP} - O_{i,t}, & \text{otherwise},
    \end{cases}
\]
where
\[
    H_{i,t:t+MHP} = \text{max}\{H_{i,t}, H_{i,t+1}, ..., H_{i,t+MHP}\}, \\
\]
\[
    L_{i,t:t+MHP} = \text{min}\{L_{i,t}, L_{i,t+1}, ..., L_{i,t+MHP}\},
\]
and $O_t$, $H_t$, $L_t$, and $C_t$ are the open, highest, lowest, and close price at day $t$.

In total, for each item in the trading universe we measure performance across 2{,}280 scenarios (MHP: 10; PT: 38; SL: 6). Given the large universe and the 10 different trading signals (horizon 1 through horizon 10 prediction signals), the computational load to extract the optimal execution values is substantial. Therefore, we use Microsoft Azure to containerize and run the computation in parallel.

To efficiently execute this large-scale scenario analysis, we leverage Microsoft Azure Kubernetes Service (AKS) for parallel computation. To manage the grid search for optimal execution parameters (MHP, PT, and SL), which requires substantial compute, we containerize the jobs and distribute the workload across 20 pods. Each pod is provisioned with 1.6 CPU cores and 6,Gi of memory, and the average runtime per pod is approximately 8 days (192 hours). The underlying infrastructure uses Standard D4s v3 virtual machines, each offering 4 vCPUs and 16,Gi RAM, providing a balanced compute environment. The AKS cluster is autoscaled to 20 nodes to accommodate the distributed jobs efficiently.

From a cost perspective, we estimate the cloud compute expenditure as follows. The Standard D4s v3 VM is priced at approximately \$0.376 per hour. Given the configuration of 10 VMs supporting 20 pods over 192 hours, total VM usage amounts to 1{,}920 hours, leading to a total compute cost of \$722 (Table \ref{table:azure_cost_table} summarizes the AKS configuration and cost). As obtaining the optimal execution values is a one-time task that does not need to be repeated during trading, using cloud infrastructure makes the scenario analysis both scalable and cost-effective.

\begin{table}[h]
    \centering
    \caption{AKS configuration and cost.}
    \label{tbl:aks}
    \begin{tabular}{@{}ll@{}}
        \toprule
        Pods                               & 20 (approximately 20 tickers per pod) \\
        Pod resources                      & 1.6 vCPU, 6 GiB RAM \\
        Node type                          & \texttt{Standard D4s v3} (4 vCPU, 16 GiB RAM) \\
        Autoscale window                   & 1 -- 20 nodes \\
        Wall-clock runtime                 & $\sim$8 days \\
        Total VM-hours                     & 1\,920 \\
        Total VMs                          & 10 \\
        Azure list price                   & \$0.376 per hour \\
        \textbf{Total compute cost}        & \textbf{\$722} \\
        \bottomrule
    \end{tabular}
    \label{table:azure_cost_table}
\end{table}

\begin{figure}
    \centering
    \includegraphics[width=\textwidth]{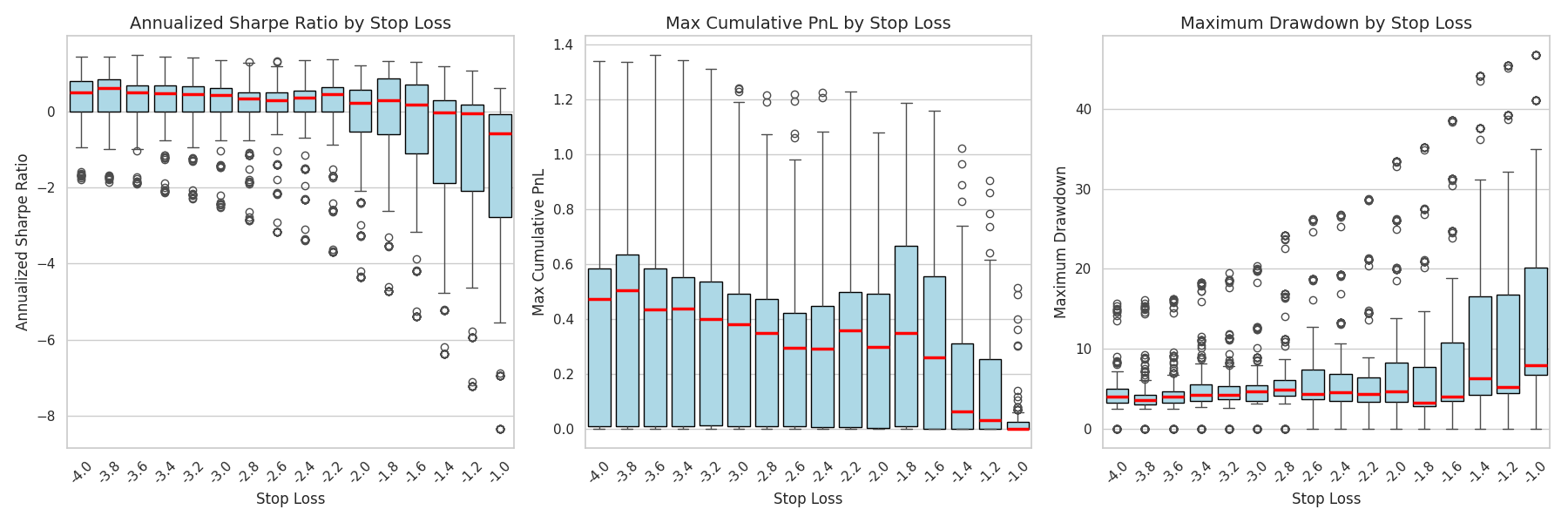}
    \caption{Boxplot showing the distribution of Sharp Ratio, cumulative return/PnL (\%), and Drawdown for different cases in the scenario analysis.}
    \label{fig:stop_loss_scenario_analysis}
\end{figure}

For each ticker-signal pair, the system produced a comprehensive set of outputs. For every scenario evaluated, three visualizations summarize performance under variations in MHP, PT, and SL thresholds. Figure \ref{fig:stop_loss_scenario_analysis} shows an example for Commvault Systems, Inc. (CVLT). The figure illustrates scenario performance across different stop-loss levels. As shown, most scenarios perform better when the Profit-Taker (PT) is set to 3.8\% of the entry price. 

Moreover, the results of each case are stored in a dataframe detailing the performance metrics. These results are saved to Azure Blob Storage and made available for further analysis. This scalable, cloud-based setup enabled a comprehensive scenario analysis, allowing us to test parameterized signal-execution strategies at scale while maintaining full reproducibility and minimizing operational overhead. 

\section{Accuracy and statistical significance}\label{sec:accuracy}

We evaluate the directional accuracy and performance of trading signals across our entire trading universe. In this section, we present the results of our analysis, including detailed accuracy scores and statistical significance of the measured accuracies.

\subsection{Accuracy}

When a signal is positive, a \textbf{long} position is opened on the next available trading day. When a signal is negative, a \textbf{short} position is opened on the next available trading day. For each asset, the directional accuracy over a specific \emph{holding period} $h$ is computed with respect to the sign of its return over that period. Consistent with the MHP search space in the previous section, we examine trade performance for 11 different holding periods (business days), labeled from 0 to 10. The multi-horizon-ahead return for each ticker is calculated by comparing the close at holding period h, i.e., day $t+h$, with the open on day $t+1$. Since positions are opened one business day after the signal is generated, we assume the signal pertains to day $t+1$.\footnote{As previously discussed, the signal is generated at the end of the trading session on day $t$ to forecast days $t+1,t+2,\dots,t+10$. This ensures any potential look-ahead bias is addressed in all stages of analysis and simulation.} For example:

\begin{itemize}
    \item A holding period of \textbf{0} indicates that the return is calculated as the percentage change between the open on the next available day after the signal is generated and the close of that same day.
    \item A holding period of \textbf{1} indicates that the return is the percentage change between the open price on the next available day and the close price one day later.
    \item Similarly, a holding period of \textbf{k} indicates that the return is the percentage change between the open on the next available day after the signal is generated and the close \textbf{k} days after the position is opened.
\end{itemize}

For each ticker, we track how often the direction of the generated signals aligns with the direction of the return for the same signal horizon and holding period. Accordingly, \textbf{accuracy} is computed as the ratio of correctly predicted directions to total predictions:
\[
    \text{Accuracy} = \frac{\text{Number of correctly predicted directions}}{\text{Total number of predictions}} \times 100\%.
\]

The final aggregated results of the accuracy analysis for 814 stocks consist of the items defined as follows:

\begin{itemize}
    \item \textbf{Ticker}: The stock ticker symbol (e.g., AAPL).
    
    \item \textbf{long\_0}, \textbf{long\_1}, \ldots, \textbf{long\_10}: The accuracy (in percentage) of all long signals held for 0, 1, \ldots, 10 days, respectively.
    
    \item \textbf{avg\_long}: The average accuracy (in percentage) of all long signals across all holding periods (from 0 to 10).
    
    \item \textbf{max\_long}: The maximum accuracy among long\_0, long\_1, \dots, long\_10.
    
    \item \textbf{min\_long}: The minimum accuracy among long\_0, long\_1, \dots, long\_10.
    
    \item \textbf{pct\_long}: The percentage of signals that were long out of all signals for the ticker.
    
    \item \textbf{best\_day\_long}: The specific holding period (0--10) at which the long strategy  achieved max\_long accuracy.

    \item \textbf{short\_0}, \textbf{short\_1}, \ldots, \textbf{short\_10}: The accuracy (in percentage) of all short signals held for 0, 1, \ldots, 10 days, respectively.
    
    \item \textbf{avg\_short}: The average accuracy of all short signals across the same holding periods (0 to 10).
    
    \item \textbf{max\_short}: The maximum accuracy among short\_0, short\_1, \dots, short\_10.
    
    \item \textbf{min\_short}: The minimum accuracy among short\_0, short\_1, \dots, short\_10.
    
    \item \textbf{pct\_short}: The percentage of signals that were short out of all signals for the ticker.
    
    \item \textbf{best\_day\_short}: The specific holding period (0--10) at which the short strategy achieved max\_short accuracy.
    
\end{itemize}

\subsection{Statistical Significance and Reliability}

Since the long/short signal distribution is inherently class-imbalanced, we apply robust statistical methods to ensure a fair comparison of signal accuracies. To determine whether the observed accuracies are statistically significant, we conduct formal tests and report p-values and confidence intervals. To evaluate reliability and significance, we consider:

\begin{itemize}
    \item Using a \textbf{binomial test} or a \textbf{z-test} for proportions to assess whether measured accuracies deviate significantly from random chance (e.g., 50\%).
    \item Constructing \textbf{confidence intervals} for each accuracy measure (e.g., a 95\% confidence interval around the estimated accuracy).
\end{itemize}

\subsubsection{Interpretation of Statistical Metrics:}

\begin{itemize}
    \item \textbf{Accuracy (\%):} 
        This is the percentage of signals whose directions are aligned with the direction of the observed return for a particular signal type (long or short) and a particular holding period. For instance, if \texttt{long\_0} is 57.0, it indicates that out of all same-day long signals generated for that ticker, 57\% were pointing to the same direction as the true return observations.

    \item \textbf{Sample Size (e.g., \texttt{n\_long}, \texttt{n\_short}):} 
        This is the total number of long or short signals generated for each ticker. If a particular ticker had 100 long signals, then the accuracy of \texttt{long\_0} is based on those 100 signals. The larger the sample size, the more confidence one can have in the overall observed accuracy.

    \item \textbf{\texttt{p\_value\_vs\_50\%} (p-value from the proportions z-test):}
        We compare each observed accuracy to a baseline (commonly 50\%) under the null hypothesis $H_0$:
        \[
          H_0: \text{True accuracy is } 50\%,
        \]
        \[
          H_1: \text{True accuracy differs from } 50\%.
        \]
        The p-value of this test will let us how likely it is to observe the measured accuracy assuming the true accuracy is 50\%. This is done using a z-test for proportions, where the standard error ($SE_{0}$) is defined as:
        \[
          SE_{0} = \sqrt{ \frac{p_0 (1 - p_0)}{n} },
        \]
        with $p_0$ the baseline proportion (e.g., 0.5) and $n$ the number of observations.
    
        The observed accuracy $\hat{p}$ is then compared to the baseline $p_0$ using the z-score:
        \[
          z = \frac{\hat{p} - p_0}{SE_{0}},
        \]
        which quantifies how many standard errors the observed proportion is from the baseline. A larger absolute z-score indicates a greater deviation between observed and expected accuracy.
    
        A smaller standard error—occurring when $n$ is large—makes the test more sensitive to differences. Consequently, the p-value, computed from the z-score, tends to decrease with larger sample sizes if the observed accuracy remains consistently above (or below) 50\%.
        
        A low p-value (e.g.,$<0.05$) suggests that the observed accuracy is statistically different from 50\% and unlikely to arise by chance. Conversely, a high p-value indicates that the observed accuracy could plausibly occur even if the signal were generated by a random binary process.
        
        \textit{Example:} Suppose the observed accuracy is 60\% on a sample of $n=200$ trades, and we test against a baseline of 50\%:
        \[
          \hat{p} = 0.6, \quad p_0 = 0.5, \quad n = 200.
        \]
        The standard error is:
        \[
          SE_{0} = \sqrt{ \frac{0.5 \times (1 - 0.5)}{200} } = \sqrt{ \frac{0.25}{200} } \approx 0.0354.
        \]
        The corresponding z-score is:
        \[
          z = \frac{0.6 - 0.5}{0.0354} \approx 2.82.
        \]
        A z-score of 2.82 corresponds to a p-value well below 0.05, indicating that the observed accuracy is significantly better than chance.

        \item \textbf{\texttt{ci\_lower}, \texttt{ci\_upper} (confidence interval for observed accuracy):}
        In addition to computing a p-value, we estimate a confidence interval (CI) around the observed accuracy $\hat{p}$, which provides a plausible range for the true accuracy based on the sample data.
        
        For a 95\% confidence level, the interval is computed using the normal approximation (Wald interval):
        \[
          CI = \hat{p} \pm z_{\alpha/2} \times SE_{\hat{p}},
        \]
        where $\hat{p} = \frac{\text{successes}}{n}$ denotes the observed accuracy, and $z_{\alpha/2}$ is the critical value from the standard normal distribution (e.g., $1.96$ for a 95\% CI). The standard error $SE_{\hat{p}}$ is given by:
        \[
          SE_{\hat{p}} = \sqrt{ \frac{\hat{p}(1 - \hat{p})}{n} }.
        \]
    
        The lower and upper bounds of the confidence interval are:
        \[
          \text{Lower} = \hat{p} - z_{\alpha/2} \times SE_{\hat{p}}, \quad
          \text{Upper} = \hat{p} + z_{\alpha/2} \times SE_{\hat{p}}.
        \]
    
        \textit{Effect of sample size:} As the sample size $n$ increases, the standard error decreases, yielding a narrower (more precise) confidence interval. Conversely, with a small sample size, the standard error is larger, making the confidence interval wider and less certain.
    
        \textit{Example:} Suppose the observed accuracy is 60\% ($\hat{p} = 0.6$), and we compute the 95\% confidence interval using the normal method:
    
        \begin{itemize}
          \item \textbf{Case A (small sample):} $n = 20$
            \[
              SE_{\hat{p}} = \sqrt{ \frac{0.6 \times 0.4}{20} } \approx 0.1095, \quad
              CI = 0.6 \pm 1.96 \times 0.1095 = [0.385, 0.815]
            \]
    
          \item \textbf{Case B (large sample):} $n = 200$
            \[
              SE_{\hat{p}} = \sqrt{ \frac{0.6 \times 0.4}{200} } \approx 0.0346, \quad
              CI = 0.6 \pm 1.96 \times 0.0346 = [0.532, 0.668]
            \]
        \end{itemize}

        As shown above, a larger sample size leads to a tighter confidence interval around the observed accuracy.

\end{itemize}

In summary, analyzing all four elements (accuracy, sample size, p-value, and confidence intervals) offers deeper insight into the reliability of the trading signals:
\begin{itemize}
    \item \textbf{Accuracy} indicates the observed performance.
    \item \textbf{Sample size} shows how many trades underlie that observed performance.
    \item \textbf{p-value} reveals whether the result is \emph{statistically} different from 50\%.
    \item \textbf{Confidence intervals} give a range of plausible values for the \emph{true} accuracy.
\end{itemize}

The results of the signal accuracy analysis is stored for each ticker where we have the observed accuracy (in percentage), the number of signals generated (sample size), the p-value from a z-test against a 50\% success rate, and the 95\% confidence interval for the true accuracy. As an example, Table \ref{table:accuracy_CVLT} shows the accuracy results for "CommVault Systems, Inc." (CVLT), one of the items in the covered universe. In this table, apart form above mentioned information, we have the strategy that represents the direction at which the ticker should be traded and period signal that shows the best signal period for the ticker. Both of these parameters are found by analysing the optimization process in section \ref{sec:scenario_analysis_and_optimization}.

\begin{table}[htbp]
\centering
\caption{CVLT signal accuracy Summary.}
\label{tab:cvlt_summary}
\begin{tabular}{ll}
\toprule
\textbf{Metric} & \textbf{Value} \\
\midrule
Ticker & CVLT \\
Name & CommVault Systems, Inc. \\
Strategy & Long Only \\
Period Signal & 3 \\
Pct\_long & 64.2 \% \\
Long Accuracy & 67.76 \% \\
Short Accuracy & 59.62 \% \\
Sample\_Size & 577 \\
p\_value\_vs\_50\% & 6.86E-20 \\
CI\_95\_lower\_\% & 63.95 \% \\
CI\_95\_upper\_\% & 71.57 \% \\
\bottomrule
\label{table:accuracy_CVLT}
\end{tabular}
\end{table}

As we can see, the occurrence of the long trading signals for CVLT is relatively high at 64.2\% of the generated signals, and its confidence interval and p-values show that the measured accuracy for this stock is statistically significant. Beyond summary statistics such as long accuracy and short accuracy, this extended evaluation allows us to statistically interpret how reliable each signal truly is. Two tickers may have identical accuracies, but one may be statistically significant (based on sample size and confidence intervals), while the other is not.

\begin{table}[htbp]
\centering
\caption{P-value Summary Statistics.}
\label{tab:p-value_summary}
\begin{tabularx}{\textwidth}{YYYYYYYYY}
\toprule
\textbf{Signal Type} & \textbf{Mean} & \textbf{Med.} & \textbf{Std. Dev.} & 
\textbf{Min} & \textbf{Max} & \textbf{\% $p<$1\%} & \textbf{\% $p<$5\%} & \textbf{\% $p<$10\%} \\
\midrule
Long  & 0.0259 & 0.0002 & 0.0670 & 0.0000 & 0.5979 & 74.948 & 86.349 & 90.900 \\
Short & 0.0367 & 0.0005 & 0.0823 & 0.0000 & 0.5742 & 68.985 & 81.642 & 86.402 \\
Both  & 0.0313 & 0.0003 & 0.0752 & 0.0000 & 0.5979 & 71.967 & 83.996 & 88.651 \\
\bottomrule
\end{tabularx}
\end{table}

Taking a broader picture of the statistical test results, the summary presented in Table \ref{tab:p-value_summary} reports the distributional characteristics of the p-values obtained across all tickers, separately for long, short, and both signals. Several important features stand out. First, the mean p-values are low (around 0.026 for long signals and 0.037 for short signals), which already suggests that, on average, the null hypothesis of random 50\% accuracy is strongly rejected. The medians are even smaller (0.0002 for longs and 0.0005 for shorts), highlighting that the majority of p-values are concentrated near zero rather than around conventional significance cutoffs.

Most importantly, the proportion columns show that the vast majority of signals meet conventional statistical significance levels. For example, more than 74\% of long-signal p-values fall below 1\%, and over 86\% fall below 5\%. Even for short signals, 69\% are below 1\% and over 81\% below 5\%. When both directions are aggregated, nearly 84\% of all signals achieve 5\% significance. Such proportions are high for financial prediction tasks, where genuine edge is typically weak, multiple-testing is pervasive, and microstructure noise and near-efficiency usually limit the share of signals clearing 5\%—let alone 1\%—especially after accounting for serial dependence and out-of-sample validation.

\begin{figure}
    \centering
    \includegraphics[width=\textwidth]{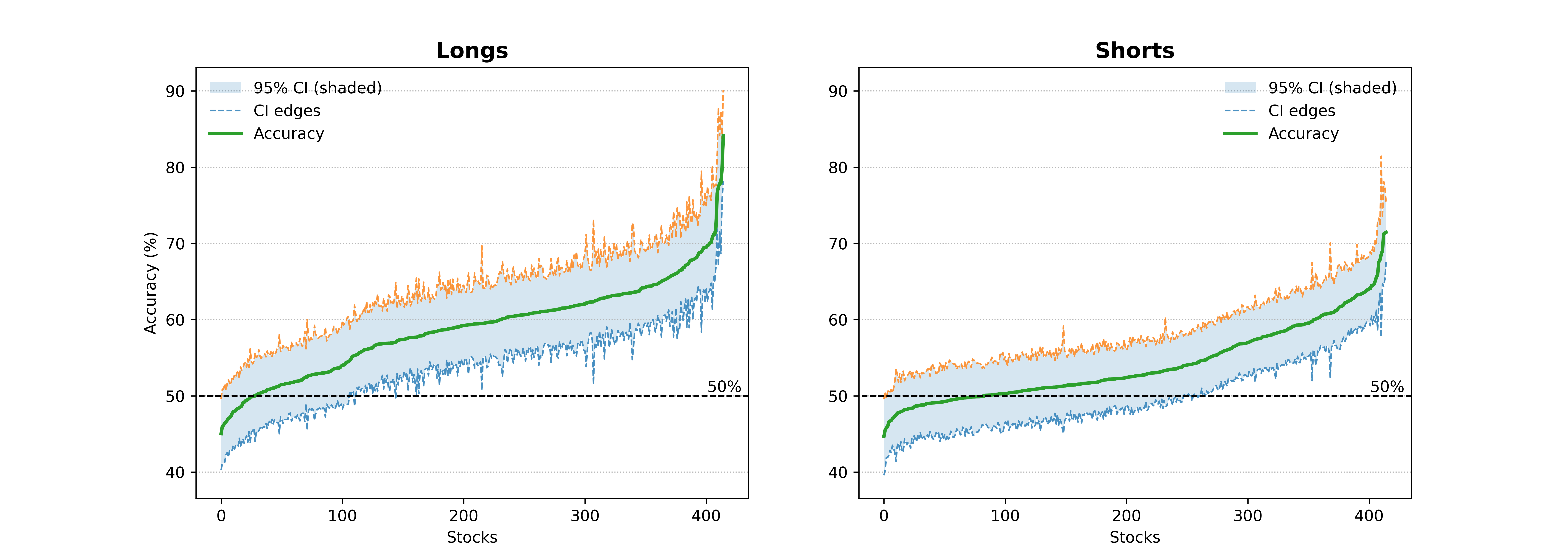}
    \caption{Confidence interval plots for the period beginning from the first observation to the end of 2024. The stocks selected from the optimization process are ordered from lowest to highest accuracy on the horizontal axis. The solid green line shows the average accuracy for each ticker across the sample and the shaded areas shows the 95\% confidence interval. The left plot shows long signals, while the right plot depicts short signals.}
    \label{fig:ci_plots_inception}
\end{figure}

\begin{figure}
    \centering
    \includegraphics[width=\textwidth]{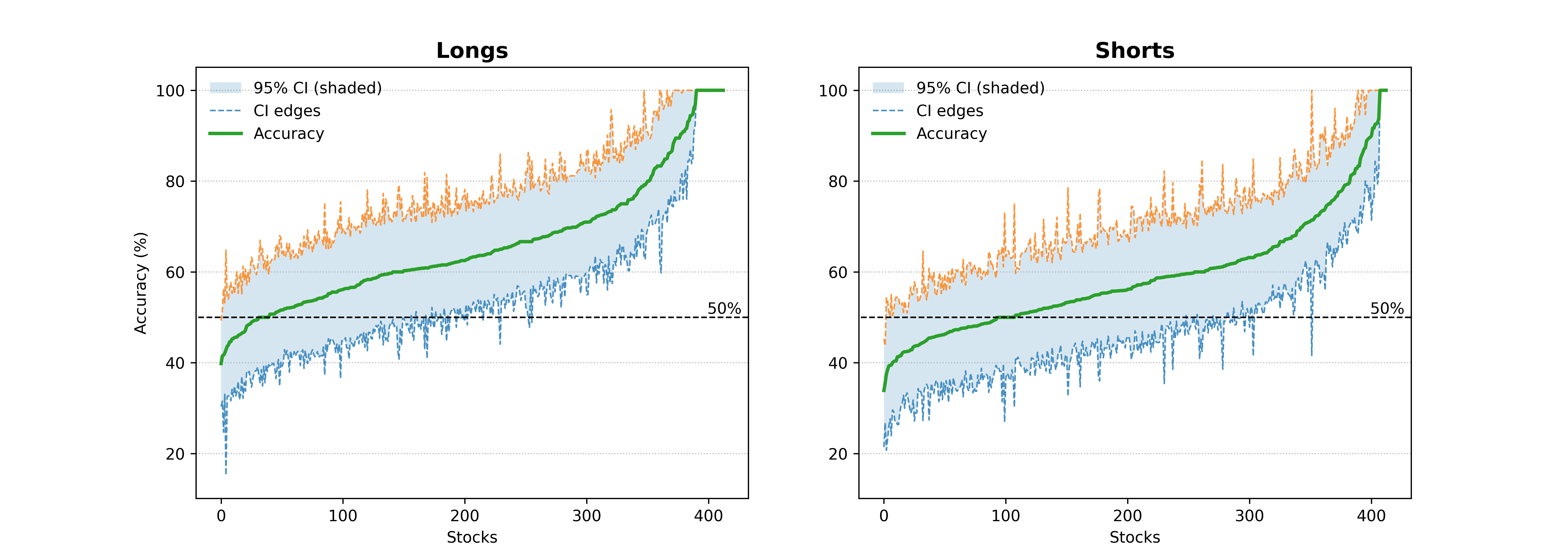}
    \caption{Confidence interval plots for the period of the first two quarters of 2025. The stocks selected from the optimization process are ordered from lowest to highest accuracy on the horizontal axis. The solid green line shows the average accuracy for each ticker across the sample and the shaded areas shows the 95\% confidence interval. The left plot shows long signals, while the right plot depicts short signals.}
    \label{fig:ci_plots_2025}
\end{figure}

Figure \ref{fig:ci_plots_inception} displays the cross-universe accuracy and 95\% confidence intervals for the long and short signals over the period from the first observation to the end of 2024, with tickers ordered by increasing observed accuracy on the horizontal axis. The solid green line shows the observed accuracy for each ticker, computed using the best-performing signal period and holding period identified for that stock; the shaded band and dashed edges show the empirical 95\% confidence interval obtained from the z-test (binomial approximation) around the observed accuracy.

Two salient features emerge. First, the distribution of observed accuracies is skewed above the 50\% benchmark: for long signals, more than 90\% of tickers achieve an accuracy above 50\%, while for short signals this figure is around 75\%. Moreover, many of these also have 95\% lower bounds above 50\%, indicating statistical evidence of predictive power beyond chance. Second, the confidence intervals maintain a relatively stable distance from the mean accuracy line across the entire ranking of tickers. This indicates that the level of statistical uncertainty is consistent regardless of whether the observed accuracy is moderate or high. In other words, the model does not become less reliable as accuracy improves; rather, the uncertainty remains controlled throughout the distribution. If the confidence intervals had widened disproportionately with higher accuracies, this would have suggested instability in the signal reliability. Instead, the uniform separation between the accuracy curve and its confidence bounds underscores the robustness of the results across the full range of signals.

Comparing panels, long signals are generally more accurate than short signals. This is consistent with the well-known upward drift in equity prices (positive equity risk premium), which creates an asymmetric environment where long-only signals are, on average, easier to exploit. However, the figure also shows many high-accuracy short signals, demonstrating that the signal generation process produces useful information on both sides of the market.

Figure \ref{fig:ci_plots_2025} shows a similar plot for the first and second quarters of 2025. The analysis follows the same procedure as above (accuracies computed using each stock’s best-performing signal period and holding period), but the 95\% confidence intervals are generally wider and their boundaries more volatile. This widening largely reflects averaging over a smaller sample observations per ticker and elevated market uncertainty during the period, both of which increase estimation variance. Despite the larger uncertainty, the model preserves its edge: a large majority of tickers remain above the 50\% benchmark, and several tickers register perfect (100\%) realized accuracy over the sample considered.

\section{Risk and Return}

While accuracy provides an initial indication of a trading signal’s potential effectiveness, it is not the sole determinant of success in financial markets. Ultimately, the most critical factors are the strategy’s returns and associated risks. To assess this balance, we rely on key performance metrics such as return, Profit and Loss (PnL)\footnote{The concepts of PnL and returns are both used to refer to trade/portfolio returns in percentages throughout this study and used interchangeably.}, Sharpe ratio (SR), and drawdown. Having evaluated the accuracy and statistical significance of the trading signals in the previous section, we now analyze the strategy’s return profile and risk characteristics using these financial metrics.

\subsection{Extracting Return and Risk Metrics}

To evaluate the strategy’s risk-return characteristics, we use outputs from the scenario analysis and optimization step in Section \ref{sec:scenario_analysis_and_optimization}, which identified the optimal Maximum Holding Period (MHP), Profit-Taker (PT), and Stop-Loss (SL) settings. These parameters, together with the OHLC price data, are used to compute both return and risk metrics for each ticker.

\paragraph{Cumulative return/PnL:} Cumulative return/PnL reflects the aggregated return of the strategy over time. It measures the percentage profit or loss generated by sequentially applying the trading signals over the historical price series, assuming no reinvestment of gains\footnote{All cumulative returns in this study are calculated using a simple, non-compounded aggregation of returns, where each period’s return is computed against a fixed notional capital base.}. This metric captures both the frequency and magnitude of successful signals and provides an intuitive view of the strategy’s performance trajectory.

\paragraph{Drawdown:} Drawdown measures the percentage decline from a historical peak in cumulative returns to a subsequent trough. It captures the impact of sequential negative returns. \textbf{Maximum drawdown (MDD)} represents the largest observed drawdown over the evaluation period and serves as a key risk metric, highlighting the strategy’s worst-case peak-to-trough loss and informing capital-preservation considerations under adverse conditions.

\paragraph{Annualized Sharpe Ratio:} The Sharpe ratio measures risk-adjusted performance by comparing the portfolio’s excess return over the risk-free rate to the volatility of its returns. To enable comparability across assets and strategies, it is typically expressed in annualized terms. A higher Sharpe ratio reflects more efficient compensation for risk undertaken.

\subsection{Visualization of Risk-Return Characteristics}

\begin{figure}
    \centering
    \includegraphics[width=\textwidth]{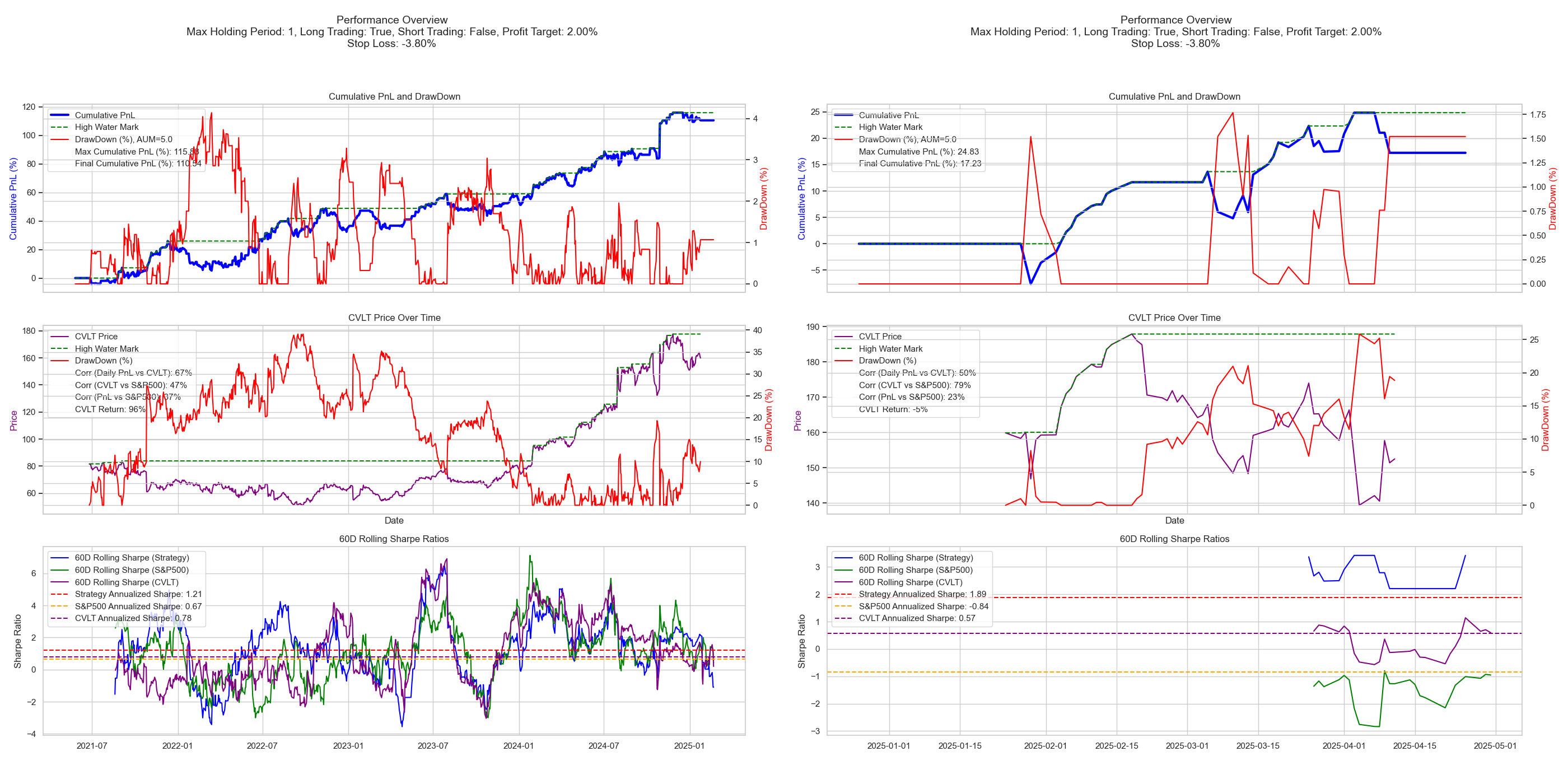}
    \caption{Visualization of the strategy’s returns and risk, compared with the macro benchmark (S\&P 500) and the stock's buy-and-hold.}
    \label{fig:return_risk}
\end{figure}

To better understand these metrics, we provide a three-part plot (Figure \ref{fig:return_risk}) visualizing the return and risk profiles of both the trading strategy and a benchmark buy-and-hold approach:

\begin{itemize}
  \item \textbf{Top Subplot:} Shows the cumulative return/PnL in percentages (left y-axis) and drawdown (right y-axis) over time, with the high watermark displayed as a dotted line.
  \item \textbf{Middle Subplot:} Displays the price of CVLT (buy-and-hold) along with its drawdown, allowing comparison with the strategy’s behavior. The legend includes correlation values between:
    \begin{itemize}
      \item Daily return/PnL and CVLT price
      \item CVLT and S\&P 500
      \item return/PnL and S\&P 500
    \end{itemize}
  \item \textbf{Bottom Subplot:} Shows the 60-day rolling Sharpe ratios of the trading strategy, CVLT buy-and-hold, and S\&P 500, with annualized Sharpe ratios shown as horizontal dashed lines.
\end{itemize}

\paragraph{Key Takeaways from the plot is:}
\begin{itemize}
  \item The cumulative return/PnL of the strategy is significantly higher than the CVLT buy-and-hold return.
  \item The strategy’s drawdown is approximately 10 times smaller than that of the CVLT buy-and-hold.
  \item The annualized Sharpe ratio of the strategy is greater than those of both CVLT and the S\&P 500, demonstrating superior risk-adjusted returns.
\end{itemize}

\subsection{Stress Testing the Strategy}

While the prior analysis focused on the period from June 2021 to January 2025, market conditions changed dramatically after January 2025.

\paragraph{Market Turbulence After January 2025:} Following accelerated interest rate hikes, geopolitical tensions, and tech sector corrections, global markets saw increased volatility. Major indices experienced drawdowns exceeding 20\%, correlations across asset classes broke down, and liquidity became strained. This period presented a substantial challenge for most trading systems.

\paragraph{Stress Test Methodology:} We segmented the data into ``Before January 2025'' and ``After January 2025'' periods. Our goal was to evaluate whether the trading signals remained robust in these drastically different regimes.

From Figure \ref{fig:return_risk}, we observe that even tough the returns of the S\&P 500 and buy-and-hold scenario were not positive, our strategy signals managed to produce a positive return. This indicates that the signal performance was stable and unchanged during the harsh market regime.

\begin{figure}
    \centering
    \includegraphics[width=0.8\textwidth]{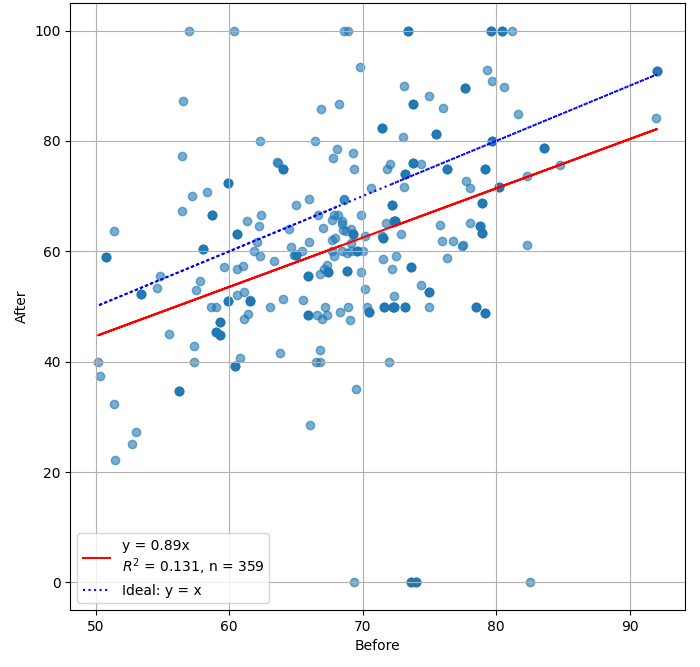}
    \caption{Scatter plot showing the long accuracy of the trading signals, before and after January 2025.}
    \label{fig:scatter_long}
\end{figure}

\begin{figure}
    \centering
    \includegraphics[width=0.8\textwidth]{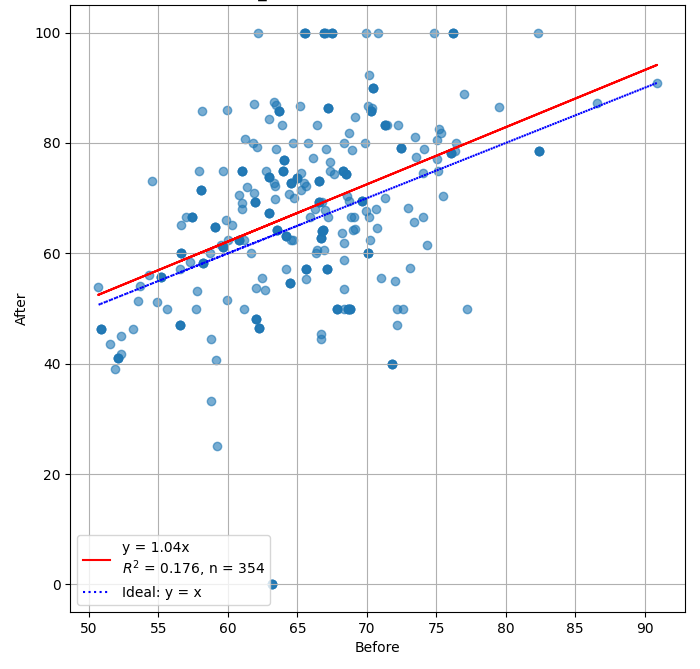}
    \caption{Scatter plot showing the short accuracy of the trading signals, before and after January 2025.}
    \label{fig:scatter_short}
\end{figure}

\paragraph{Results:} The scatter plots (see Figure \ref{fig:scatter_long} and Figure \ref{fig:scatter_short}) display ``Before'' vs ``After'' accuracy of our sample. 

Key findings include:
\begin{itemize}
  \item The regression lines are noticeably close to the ideal $y = x$ line, indicating stability in signal performance across regimes.
  \item The slope and $R^2$ values demonstrate consistency and predictive strength.
\end{itemize}

\paragraph{Conclusion:} These findings strongly support that:
\begin{itemize}
  \item The trading strategy is robust and largely uncorrelated with broader market movements.
  \item It continues to function effectively during historically adverse periods.
  \item The low correlation to CVLT and the S\&P 500 implies diversification benefits.
\end{itemize}

\section{Portfolio Construction and Dynamic Rebalancing}
\label{sec:portfolio}
In the previous section, we evaluated the generated signals’ ability to predict the price movement of individual stocks. In what follows, we translate these insights into a practical investment strategy with an example of systematic portfolio construction and dynamic rebalancing. We detail the methodology employed to build and adapt a trading portfolio using our generated signals and evaluate the portfolio’s performance over time. We adopt a \emph{walk-forward} design that strictly separates model evaluation from live decision-making.

The portfolio construction follows a six-quarter rolling window with one-quarter steps. More specifically, we start by using model outputs between mid-2021 and the end of 2022, corresponding to six quarters, as a calibration window. During this period, we aggregate daily directional signals across 814 U.S. equities and compute performance metrics—including directional accuracy, annualized Sharpe ratio, maximum drawdown, etc.—for each stock. For the stock-selection process, we sort the stocks based on their performance in each rolling selection window and select the top-performing names.

The constructed portfolio is then traded over the subsequent quarter, specifically from the beginning of 2023 through the end of Q1 2023, using the latest daily predictions from the model. At the end of each quarter, the portfolio is rebalanced: the prior 18 months of trading signals are analyzed, and new top-performing stocks are selected to form the next quarter’s portfolio. This rolling mechanism ensures that the portfolio adapts to evolving market conditions while maintaining a systematic, data-driven foundation. Note that the selection of a stock does not guarantee its inclusion in the portfolio; this is a watch list, and stocks are included only if the signals generated for that stock during the trading quarter—e.g., the first quarter of 2023—are in line with the desired trading direction assigned to the stock as a result of the optimization process. However, based on the analysis in the results section, the majority of the selected stocks are traded during the trading quarter.

We continue this rebalancing process through the end of 2024 and into the first and second quarters of 2025, resulting in a comprehensive evaluation of the dynamic portfolio strategy across multiple market regimes. Therefore, this example demonstrates the practical viability and robustness of signals generated by our model in a real-world trading context.

Our framework is inherently robust to survivorship bias due to the way the trading universe and backtest design are structured. Since mid-2021, the model has continuously generated daily predictions for a dynamic universe of over 800 U.S. equities. When a company was delisted or merged, signal generation for that ticker was automatically discontinued, and its historical signals remained archived to preserve a complete and immutable record. For performance evaluation, we restricted backtests to stocks with uninterrupted signal histories over the full study horizon, ensuring temporal consistency rather than selective inclusion. This approach eliminates the upward bias that can arise when only surviving stocks are considered. Moreover, because the live inference system produces timestamped, forward-looking predictions in real time, without retroactive adjustment, every security’s lifecycle is captured faithfully. In addition, our dynamic, walk-forward portfolio construction further mitigates survivorship bias by re-evaluating the investable universe every quarter and selecting portfolios based solely on the most recent six-quarter performance window. As a result, inclusion in or exclusion from the trading portfolio is determined adaptively rather than retrospectively, preventing the inadvertent advantage that would otherwise favor stocks that happened to persist throughout the full sample. The combination of dynamic universe management, complete archival retention, and a six-quarter consistency requirement for portfolio inclusion collectively ensures that our performance metrics reflect genuine predictive strength rather than survivorship effects.

\subsection{Results}

This section reports the out-of-sample trading results, examining how the strategy built from the model’s signals compares with a passive S\&P 500 buy-and-hold benchmark. Because the principal aim is to demonstrate the economic value of the signals, our focal strategy is the maximum drawdown-based, equally weighted portfolio. Prior to selection, we remove stocks with too few observations over the selection window and exclude stocks with market beta greater than one. We then rank the remaining universe according to the chosen metric: for risk-oriented measures such as maximum drawdown (MDD), stocks are ordered from lowest to highest and the lowest-MDD names are selected; for return-oriented measures such as the Sharpe ratio (SR), stocks are ordered from highest to lowest. For the MDD-based portfolio, at each rebalancing quarter we take the 20 lowest-MDD names from the stocks selected for long trading during the optimization process and the 20 lowest-MDD names from the stocks selected for short selling. We compare cumulative returns, risk-adjusted performance, drawdown behavior, and time-varying performance of this portfolio to the passive benchmark. The effects of alternative selection criteria are examined later in the section. Linearly weighted variants deliver qualitatively similar behavior with slightly weaker outcomes; where relevant, we summarize those results but do not plot them. All performance statistics are computed on the rolling, walk-forward trading sample described in Section \ref{sec:portfolio}.

\begin{figure}
    \centering
    \includegraphics[trim=5mm 15mm 5mm 35mm,clip,width=\textwidth]{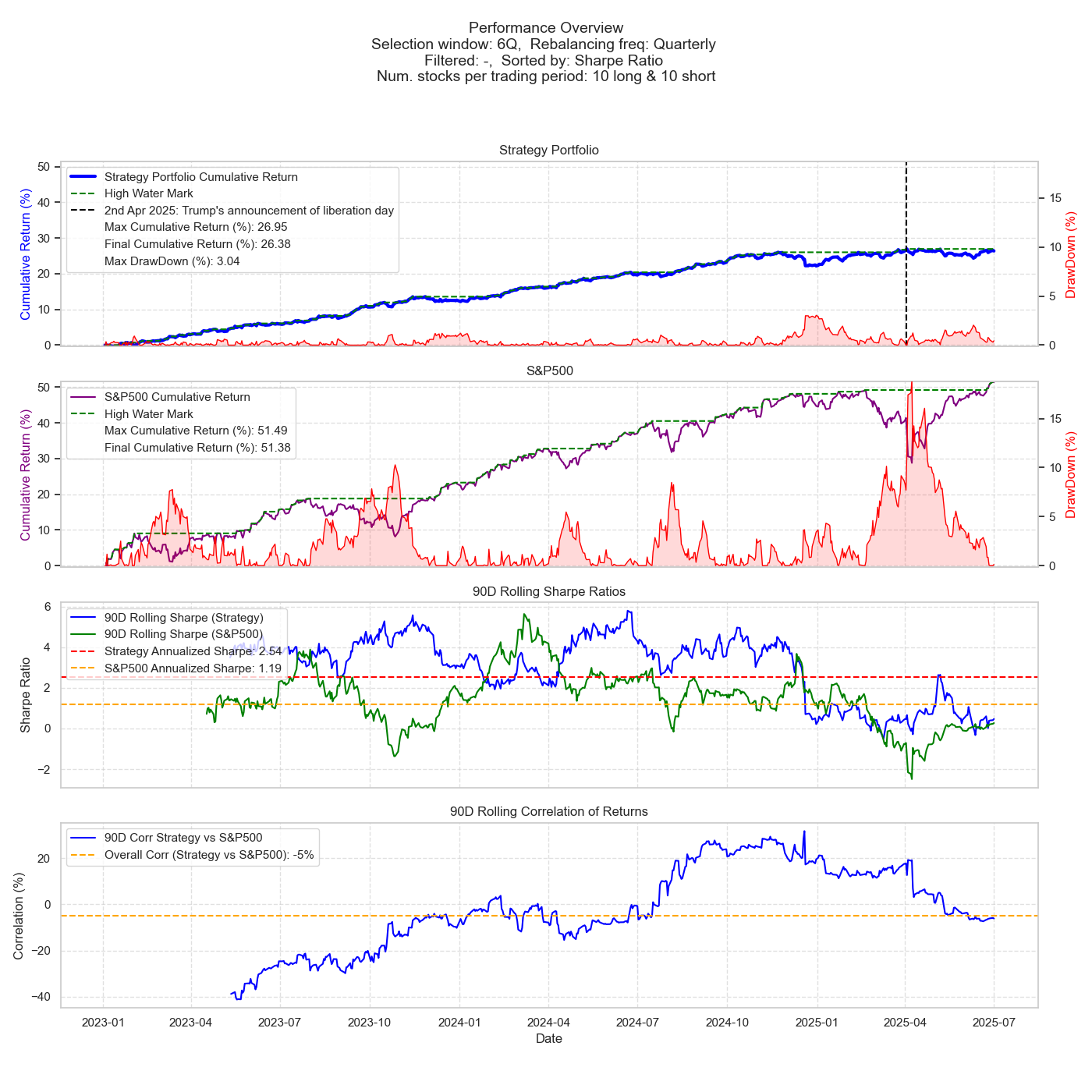}
    \caption{Out-of-sample performance of the \textbf{MDD-based, equally-weighted (MDDEW)} portfolio, benchmarked against the passive S\&P 500 buy-and-hold.}
    \label{fig:MDDEW portfolio}
\end{figure}

\subsubsection{Portfolio performance}

\paragraph{\textbf{Aggregate returns and risk-adjusted performance:}} Figure \ref{fig:MDDEW portfolio} depicts the out-of-sample performance of the MDD-based, equally weighted (MDDEW) portfolio, benchmarked against a passive S\&P 500 buy-and-hold. Over the test interval, the strategy’s final cumulative return is $\approx +26.4\%$ (peak cumulative return $\approx 26.95\%$) with a maximum drawdown of $\approx 3.0\%$ (values read from the annotated figure). Although the passive S\&P 500 over the same interval delivers larger nominal appreciation (final cumulative return $\approx +51.4\%$), the MDDEW portfolio exhibits higher risk-adjusted performance. The strategy’s annualized Sharpe ratio (computed from daily returns and displayed in the plot legend) is $\approx 2.54$, compared with $\approx 1.19$ for the S\&P 500. Thus, the MDDEW portfolio achieves higher return per unit of daily volatility despite producing a lower absolute cumulative return. This difference—smaller nominal gains but higher Sharpe—summarizes the main empirical result: the model’s signals can be converted into a tradable book that is more efficient on a risk-adjusted basis than passive market exposure.

\paragraph{\textbf{Correlation with market returns:}} The relation between the strategy and broad market returns is small on average and time varying. The annotated overall correlation between the strategy’s daily returns and S\&P500 returns is $\approx$ -5\%, indicating that MDDEW portfolio is largely orthogonal to the benchmark over the full sample. This low average correlation implies that, from a portfolio construction standpoint, MDDEW portfolio behaves like an idiosyncratic (alpha) sleeve that can complement passive equity exposure; it does not appear to be a levered or variant expression of market beta.

The figure’s 90-day rolling correlation series, however, reveals that this correlation is not constant: correlation drifts upwards through 2024—reaching positive territory in mid-2024—before falling back toward slightly negative values in 2025. That time dependence indicates that MDDEW portfolio’s diversification benefit is regime dependent: in some subperiods the strategy’s returns becomes more aligned with market moves (reducing diversification), while in others it provides a defensive, low-correlation source of returns. Any practical allocation that pairs MDDEW portfolio with passive holdings should therefore treat correlation as a variable quantity and consider dynamic sizing or stress testing for periods of elevated co-movement.

\paragraph{\textbf{Drawdowns and resilience through market events:}} A comparison of the drawdown traces highlights the different risk profiles of MDDEW portfolio and the S\&P 500. The S\&P line shows larger and deeper drawdown episodes (notably the drawdown spike visible in the middle panel), whereas MDDEW portfolio’s drawdown envelope is shallow and contained (maximum drawdown $\approx$ 3\%). Importantly, MDDEW portfolio’s limited drawdowns imply that its positive cumulative return is achieved with relatively little depth of capital erosion between highs—a property attractive to risk-conscious investors even where absolute return is lower than the benchmark.

The annotated vertical event on the plot coincides with a period of elevated market stress. The S\&P exhibits a sharp drawdown and rapid recovery during that interval, while MDDEW portfolio experiences a modest and brief deterioration in value before re-attaining prior highs. This behaviour suggests that the signals underlying MDDEW portfolio capture cross-sectional opportunities that are not tightly coupled to the market shock that produced the benchmark’s large drawdown, reinforcing the earlier observation that MDDEW portfolio is not simply capturing market beta.

\paragraph{\textbf{Time-varying behaviour -- rolling Sharpe and stability:}} The 90-day rolling Sharpe panel demonstrates that MDDEW portfolio’s performance pattern is both persistent and episodic. MDDEW portfolio’s rolling Sharpe spends the majority of the window above the S\&P’s rolling Sharpe and repeatedly attains values in the region of 2--5 during favourable subperiods, consistent with the high aggregate Sharpe ($\approx$ 2.54). This episodic concentration of high short-window Sharpe indicates that the model does not produce uniform small gains every day; instead it identifies pockets of cross-sectional opportunity that, when harvested repeatedly, build a high-quality return stream.

Toward late 2024 and into 2025 the strategy’s short-window Sharpe declines toward more moderate levels, mirroring the reduction in rolling correlation described above. The joint movement of rolling Sharpe and rolling correlation suggests a plausible narrative: when market conditions produce clear, coherent cross-sectional opportunities (for example, mid-2023 through mid-2024 in the figure), MDDEW portfolio tends to align with those trends, leading to higher gains during sustained market rallies. By contrast, in periods when the broad market weakens or fails to produce strong returns, MDDEW portfolio’s returns decouple from the benchmark—the strategy exhibits lower correlation and does not systematically amplify market losses—so that drawdowns remain contained. This dynamic — participation in market up-moves together with reduced comovement during market stress — explains why MDDEW portfolio achieves higher gains in strong market advances while helping to mitigate losses during market declines.

\paragraph{\textbf{Decomposed performance:}} Table \ref{tab:port_performance_decomp} distills the out-of-sample performance by sleeve (long, short) and in aggregate. The combined long-short book produces a return/PnL stream: cumulative return $\approx 26.38\%$, annualized Sharpe $\approx 2.54$, and an MDD of $\approx 3.04\%$. Both sides contribute—longs at $\approx 15.75\%$ (Sharpe $\approx 1.87$; MDD $\approx 3.99\%$) and shorts at $\approx 10.63\%$ (Sharpe $\approx 1.00$; MDD $\approx 3.33\%$)—highlighting persistent edge on each sleeve. Execution is nimble: mean holding periods cluster around one trading day (0.98-1.08) under a six-day realized MHP, and the program executed 8{,}859 trades (4{,}336 long; 4{,}523 short). Hit rates are consistently above 55\% (long 57.9\%, short 55.4\%, aggregate 56.6\%), in line with the signal-accuracy results and supportive of the steady upward drift in cumulative return/PnL. In aggregate, the table shows a high-Sharpe portfolio with limited drawdowns and balanced contributions from long and short exposures.

\begin{table}[htbp]
\centering
\caption{Portfolio performance metrics decomposed into long and short legs.}
\label{tab:port_performance_decomp}
{
\begin{tabularx}{\textwidth}{YYYYYYYY}
\toprule
         & \textbf{Cum. Ret./PnL (\%)} & \textbf{SR} & \textbf{MDD} & \textbf{Mean realized holding period (days)} & \textbf{Max. realized holding period (days)} & \textbf{Win rate (\%)} & \textbf{Num. transactions} \\
\midrule
Long     & 15.75                    & 1.872       & 3.988        & 1.080                               & 6                            & 57.9                   & 4336                       \\
Short    & 10.63                    & 1.001       & 3.333        & 0.885                               & 6                            & 55.4                   & 4523                       \\
Combined & 26.38                    & 2.538       & 3.039        & 0.980                               & 6                            & 56.6                   & 8859                      \\
\bottomrule
\end{tabularx}
}
\end{table}

\paragraph{\textbf{Portfolio weighting:}} Throughout, all portfolios and model settings above are implemented as equally weighted baskets. As a robustness check, we also consider a simple linearly decaying scheme in which the top-ranked names receive higher weights that decrease in a linear fashion down the ranking list. Comparing with the best model identified above (MDDEW), this tilt modestly reduces overall performance relative to equal weights. The linearly weighted variant delivers a strategy Sharpe of $\approx$ 2.25, a maximum drawdown of $\approx$ 3.37\%, a slightly negative market correlation ($\approx -4\%$), and a final cumulative return of $\approx$ 26.8\%.

\paragraph{\textbf{Role of selection metric:}} To further assess the robustness of the signals, Table \ref{tab:port_selection_summary} reports the out-of-sample performance of a broader set of portfolio selection rules and simple filters rather than focusing on a single construction. Every row in the table uses the same 6-quarter calibration window and the same quarterly rebalancing; what changes are the criterion used to rank names (Sharpe, drawdown, final cumulative return, Sortino, beta, downside risk, accuracy, etc.), the application of a simple beta filter, and the number of stocks selected. The purpose of this exercise is straightforward: we want to show how alternative, plausible selection choices affect realized returns, risk, and correlation to the market, and to assess whether the model’s signals produce performance across several settings.

The results are informative and largely consistent. Risk-aware ranking rules — in particular realized Sharpe and low-drawdown ranking — produce higher risk-adjusted outcomes. For example, ranking by Sharpe (row 1) yields a strategy Sharpe $\approx$ 1.98, a final cumulative return $\approx$ 30.7\%, and a maximum drawdown of $\approx$ 6.2\%; ranking by drawdown (row 2) gives a slightly higher Sharpe ($\approx$ 2.20) with an especially small maximum drawdown ($\approx$ 3.1\%) and a final return $\approx$ 27.3\%. Ranking by final cumulative return (row 3) and Sortino (row 4) also deliver positive cumulative returns ($\approx$ 24--31\%) and Sharpe ratios above 1.5. By contrast, naive or purely direction-based rules perform poorly: the accuracy-ranked portfolio (row 11) posts a negative final return ($\approx$ -7.8\%), a negative Sharpe ($\approx$ -0.54) and a large drawdown, underscoring that directional accuracy alone does not reliably generate tradable, risk-adjusted alpha.

Simple filters and settings alter outcomes in expected ways. Excluding high-beta names (or, conversely, requiring beta > 1) changes drawdown and return characteristics but does not erase the overall pattern: for example, applying a beta>1 filter and ranking by drawdown (row 8) produces a strong risk profiles in the table (Sharpe $\approx$ 2.38, MDD $\approx$ 2.5\%, final return $\approx$ 29.3\%), while retaining a top-10 selection under the same filter (row 7) yields comparable returns with a modestly higher drawdown. Increasing the number of selected names (row 9, top-20) slightly reduces final cumulative return but preserves a high Sharpe, illustrating that the signals remain useful when the portfolio is broadened. Two alternative ranking choices — ranking by beta alone (row 5) and ranking by downside risk (row 6) — produce weaker outcomes (low or negative Sharpe and small or negative cumulative return), which highlights that not all ex-post metrics are equally informative.

\begin{table}[htbp]
\centering
\caption{Summary of the other portfolio selection criteria and their performance.}
\label{tab:port_selection_summary}
{
\renewcommand{\arraystretch}{1.6}
\begin{tabularx}{\textwidth}{YYYYYYYYY}
\toprule
   & \textbf{Window} & \textbf{Filter out}                   & \textbf{Ranked by} & \textbf{Top n} & \textbf{SR} & \textbf{MDD} & \textbf{Corr} & \textbf{Final Cum. Ret.} \\
\midrule
1  & 6Q              & -                                     & SR                 & 10             & 1.98        & 6.23         & 9             & 30.72                    \\
2  & 6Q              & -                                     & DD                 & 10             & 2.20        & 3.14         & -1            & 27.31                    \\
3  & 6Q              & -                                     & Final Cum. Ret.    & 10             & 1.75        & 6.81         & 11            & 30.74                    \\
4  & 6Q              & -                                     & Sortino            & 10             & 1.60        & 7.36         & 11            & 24.34                    \\
5  & 6Q              & -                                     & Beta               & 10             & 0.70        & 6.70         & 6             & 7.92                     \\
6  & 6Q              & -                                     & Downside Risk      & 10             & -0.07       & 11.30        & 7             & -1.18                    \\
7  & 6Q              & Beta \textgreater 1                   & SR                 & 10             & 1.85        & 4.74         & -3            & 27.66                    \\
8  & 6Q              & Beta \textgreater 1                   & DD                 & 10             & 2.38        & 2.50         & 2             & 29.33                    \\
9  & 6Q              & Beta \textgreater 1                   & SR                 & 20             & 2.06        & 3.64         & -1            & 24.28                    \\
10 & 6Q              & Beta \textgreater 1 \& SR \textless 1 & DD                 & 10             & 1.77        & 4.53         & -1            & 24.58                     \\
11 & 6Q              & -                                     & Accuracy           & 10             & -0.54        & 78.6         & 8            & -7.77                     \\
\bottomrule
\end{tabularx}
}
\end{table}


\subsubsection{Stock turnover} \hfill \\

\begin{figure}
    \centering
    \includegraphics[width=\textwidth]{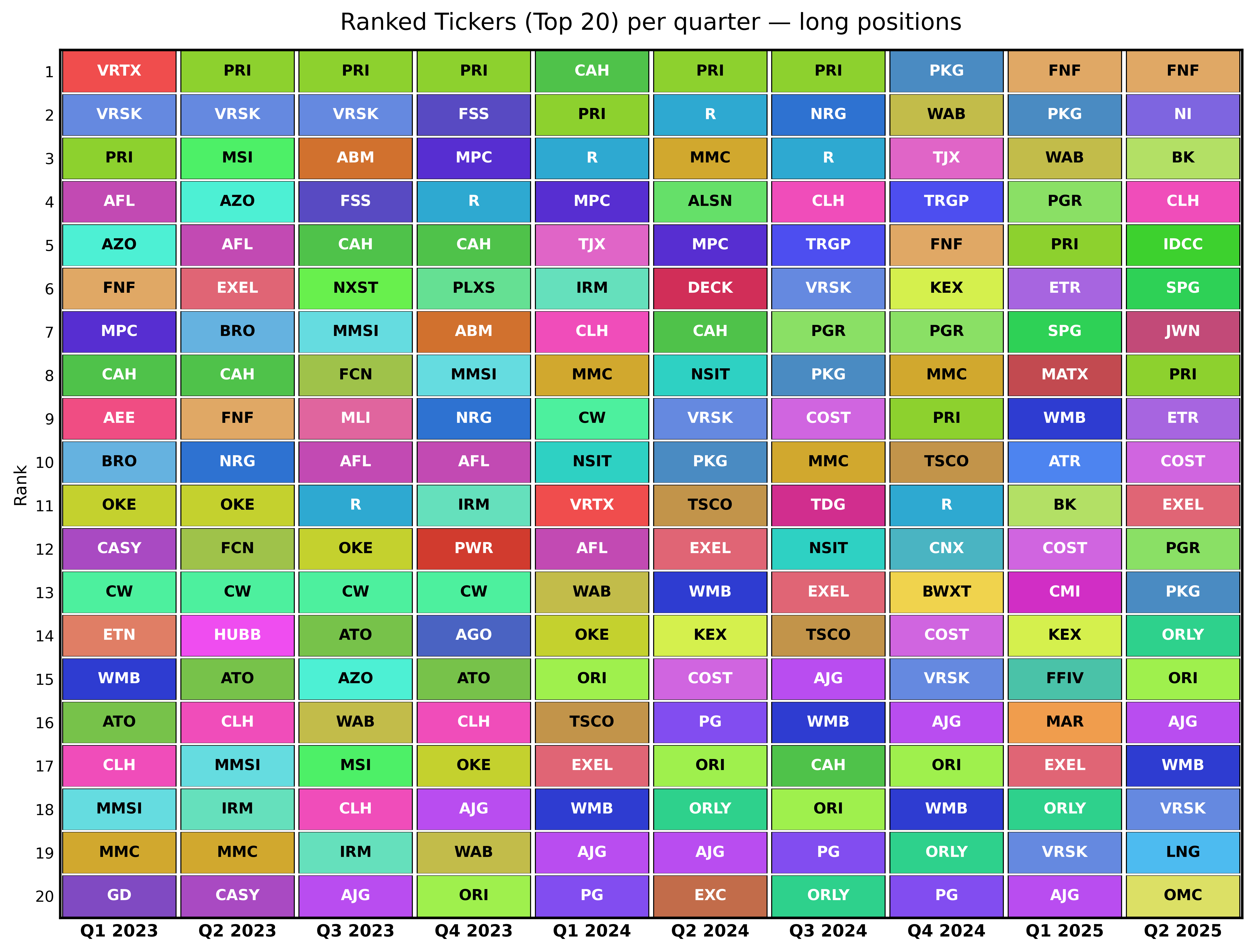}
    \caption{Color-coded quarter-by-quarter composition of the \textbf{long-leg} 20-stock trading book, displayed in a compact rank-ordered grid. Each column corresponds to a quarter (as indicated in the column header) and shows the stocks selected for long positions during that period. Rows indicate the ranking of stocks based on their historical MDD.}
    \label{fig:long_turnover}
\end{figure}

\begin{figure}
    \centering
    \includegraphics[width=\textwidth]{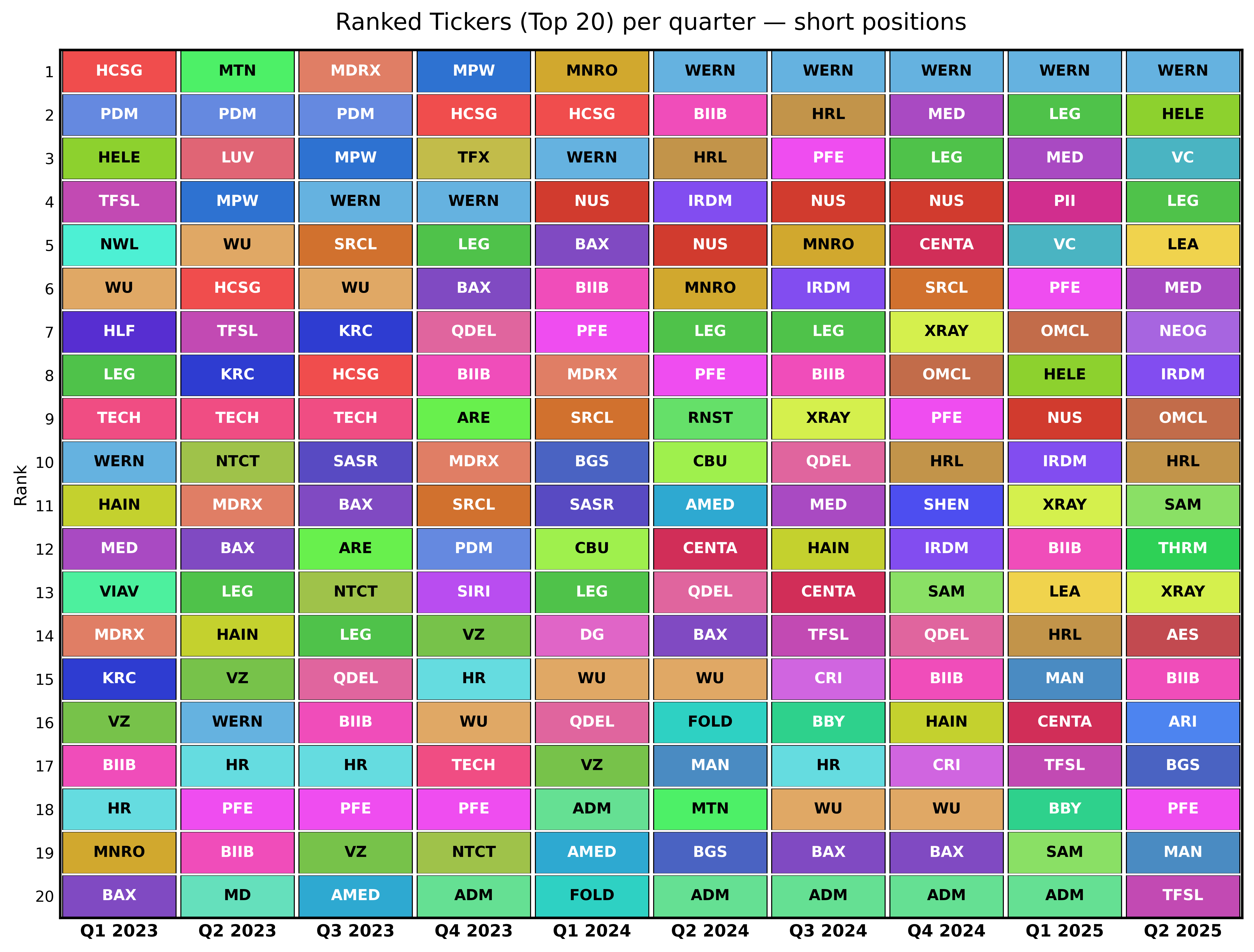}
    \caption{Color-coded quarter-by-quarter composition of the \textbf{short-leg} 20-stock trading book, displayed in a compact rank-ordered grid. Each column corresponds to a quarter (as indicated in the column header) and shows the stocks selected for long positions during that period. Rows indicate the ranking of stocks based on their historical MDD.}
    \label{fig:short_turnover}
\end{figure}

Figures \ref{fig:long_turnover} and \ref{fig:short_turnover} display the color-coded quarter-by-quarter composition of the 20-stock trading book per side in a compact, rank-ordered grid: Figure \ref{fig:long_turnover} shows the 20 names selected for long exposure in each quarter (columns = quarters, rows = rank 1..20) and Figure \ref{fig:short_turnover} shows the 20 names selected for short exposure. The visualisation is simple and categorical (one colour block per ticker) so that two structural features of the selection process are apparent: (i) how persistent particular tickers are across consecutive rebalances; and (ii) how much turnover and rank reordering occur from quarter to quarter. The following analysis describes those features and discusses their implications for implementation, risk, and interpretation of the signals.

First, the plots reveal a mixed pattern of persistence and turnover. Several tickers appear in the same side of the book in multiple consecutive quarters (for example, a handful of names in the short-panel — such as PDM and HCSG in the early quarters of the sample — recur repeatedly), which indicates that the selection rule sometimes identifies stocks with lived, stable favourable calibration metrics across the 18-month look-back. At the same time, many slots are filled by names that appear only for a single quarter before being replaced. This alternation between repeat selections and rapid replacement is consistent with a selection signal that is both stable enough to capture persistent cross-sectional patterns and responsive to regime or firm-specific changes. For a practitioner, this mix is useful: persistent names provide the basis for multi-quarter exposures (lower turnover, clearer implementation plan), while rotating names capture freshly emergent opportunities identified by the model.

Second, rank mobility within a quarter-block is informative. The matrix rows reflect the rank ordering of chosen names; in many cases the same name migrates up or down the rank ladder across adjacent quarters rather than disappearing entirely. This graded movement suggests the model is providing a continuous score (or at least a rankable quantity) rather than a binary “in/out” flag, and it implies that weighting schemes which respect rank (for example, size or linear weighting) will change realised exposures relative to simple equal-weighting. In practice, this behaviour argues for careful consideration of weighting: an aggressive weighting on rank will amplify the impact of a stock that remains ranked first across quarters, increasing concentration risk; conversely, equal weighting will reduce this effect and reduce idiosyncratic exposure from persistent top-ranked names.

Third, comparing the long and short matrices highlights asymmetries in selection dynamics. The long-side grid in Figure \ref{fig:long_turnover} shows a different pattern of reuse and rotation than the short-side grid in Figure \ref{fig:short_turnover}. Concretely, one side may contain more recurring names while the other rotates more rapidly; this asymmetry has two implications. On the modeling side it suggests that the information content of signals for “longable” names differs from that for “shortable” names (for example, some firms consistently show the metric the model rewards while others flip sign more often). On the implementation side it implies that turnover—and therefore transaction costs—will differ across the two sides of the book. Any realistic P\&L projection should therefore account for asymmetric execution friction between longs and shorts.

Finally, the plots implicitly inform concentration and diversification. Visual inspection shows periods in which the rank palette is relatively homogeneous (many different names, low persistence) and other periods in which a small set of names repeatedly occupies top ranks. Those concentrated periods raise the risk of idiosyncratic shocks: if the strategy’s gains depend heavily on a small subset of names that remain in the portfolio across rebalances, the strategy becomes exposed to single-name tail risk. Conversely, the more diversified quarters—where each column contains many different tickers—offer stronger mechanical diversification and should be less vulnerable to idiosyncratic failures. This observation reinforces the need for explicit concentration controls (position caps, volatility scaling) when moving from a simulated equally-weighted construction to a live implementation.

In sum, the colour-coded rank matrices convey two central characteristics of the selection mechanism: the model delivers both repeatable (multi-quarter) signals for a subset of names and an ability to adapt and rotate the book when prior winners decay. Those characteristics are desirable for a tradable signal set, but they also create trade-offs—between exploitation of persistent edges and protection against single-name concentration—that need to be explicitly managed in a real implementation.

\subsection{Further Discussion}

\subsubsection{The Impact of Leverage} \hfill \\

To explore the strategy’s potential when capital is not a limiting constraint, we analyze the performance of the MDDEW portfolio with a 2x leverage multiplier, assuming a 4\% annual cost of capital. This exercise sheds light on how the strategy’s core properties—its Sharpe ratio, low market correlation, and drawdown behavior—scale under higher nominal exposure.

With a 2x leverage factor, the MDDEW portfolio's cumulative return at the end of the sample is approximately +47\%, with a peak gain near +49\%. This represents an increase in absolute nominal returns compared to the non-leveraged MDDEW portfolio's cumulative return of +26.38\%. While a larger capital base and increased exposure naturally lead to higher nominal returns, the leveraged portfolio maintains an annualized Sharpe ratio of 2.26. This value, while slightly lower than the non-leveraged portfolio's Sharpe of 2.54, still exceeds the S\&P 500 (1.19). This result reinforces the conclusion from our primary analysis: the signals generate a tradable return stream whose volatility profile is attractive enough to sustain favorable risk-adjusted outcomes even under leverage.

The leveraged portfolio's daily return remains weakly correlated with the broad market, with a correlation of -5\% to the S\&P 500. This low correlation is consistent with the non-leveraged results and demonstrates that the portfolio's gains are still largely idiosyncratic and not simply an amplified bet on systematic market drivers. The time-varying performance, as shown by the rolling Sharpe series, similarly indicates that the leveraged strategy continues to produce robust, positive performance episodes throughout the test period, further validating the stability of the model’s signals.

The analysis of drawdowns reveals a key insight into the strategy's risk profile under leverage. The maximum drawdown for the leveraged portfolio increases to 5.33\% from the non-leveraged version's 3.04\%. This increase in absolute drawdown is an expected consequence of amplifying both positive and negative returns and including borrowing cost. However, the portfolio recovered and resumed its prior trajectory after these spikes, highlighting its ability to withstand adverse market events and resume profitable trading.

\begin{figure}
    \centering
    \includegraphics[trim=5mm 15mm 5mm 35mm,clip,width=\textwidth]{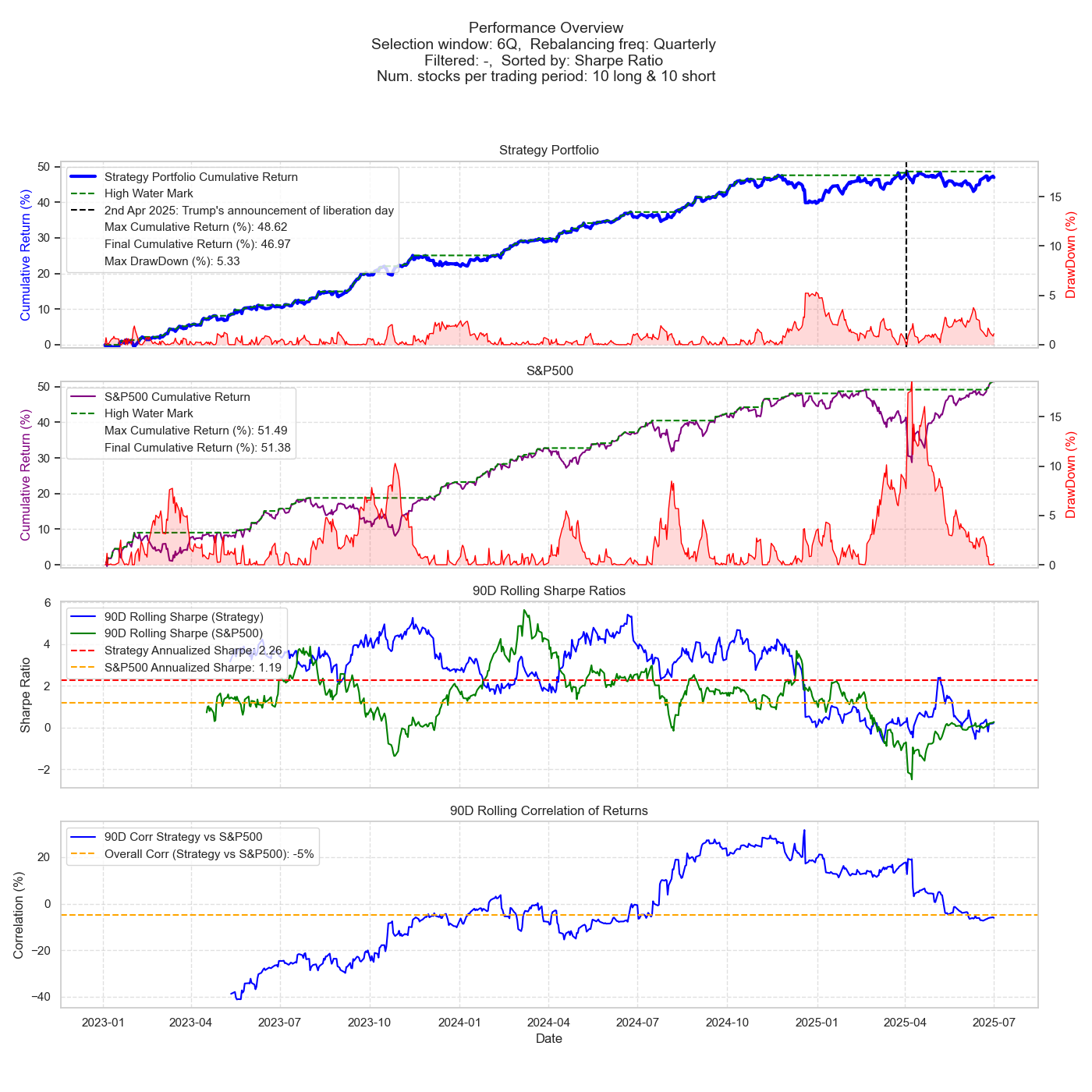}
    \caption{Out-of-sample performance of the MDD-based, equally-weighted (MDDEW) portfolio with \textbf{2x leverage and 4\% cost of capital}, benchmarked against a passive S\&P 500 buy-and-hold.}
    \label{fig:AE}
\end{figure}

\subsubsection{Beyond US equity market} \hfill \\

The current framework has been validated on a universe of large and mid-cap U.S. equities, but its architecture and feature set are naturally extendable to other markets and instruments. Within equities, one immediate extension is to small-cap stocks in the U.S., where microstructural inefficiencies may amplify the predictive power of the signals. Beyond the U.S., the same methodology can be applied to mature exchanges such as the London Stock Exchange or the Tokyo Stock Exchange, as well as selected European and emerging markets that share structural similarities with U.S. equity markets. Because the feature definitions are rooted in economically interpretable variables rather than market-specific artifacts, the signals are expected to retain predictive power across geographies. A further line of extension is to fixed-income markets. The same predictive framework that currently forecasts directional equity returns can be repurposed to model corporate bond yields, credit spreads, and rate dynamics, thereby opening an avenue into substantially larger markets. In addition, the multi-horizon design of the signals—where predictions are generated across several future time steps—lends itself to derivative markets. In particular, the short-dated horizons align with the maturities of futures contracts and short-lived options, including zero-day-to-expiry (0DTE) options that have recently gained significant liquidity. This adaptability suggests that the proposed framework is not confined to equity returns alone but can evolve into a more general platform for multi-asset signal generation and risk forecasting across global capital markets.

\subsubsection{Updated results - Q3 2025 extension} \hfill \\

\begin{figure}
    \centering
    \includegraphics[trim=5mm 15mm 5mm 35mm,clip,width=\textwidth]{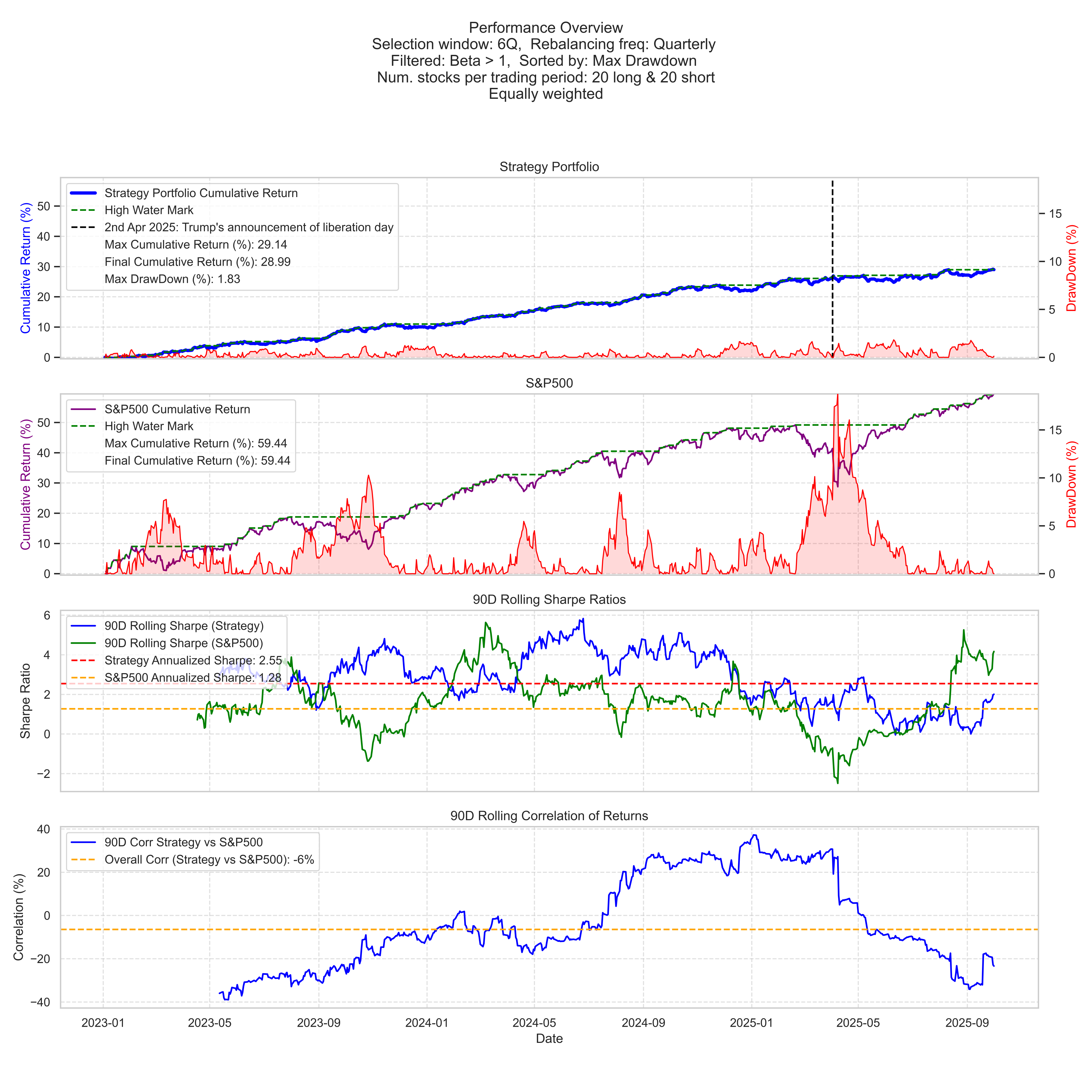}
    \caption{Out-of-sample performance of the MDD-based, equally-weighted (MDDEW) portfolio extended up to \textbf{Q3 2025}, benchmarked against a passive S\&P 500 buy-and-hold.}
    \label{fig:Q3_Non_lev}
\end{figure}

Extending the backtest through Q3 2025 confirms the continued stability of the MDDEW strategy. As shown in Figure \ref{fig:Q3_Non_lev}, the non-leveraged portfolio maintains a smooth equity curve with a final cumulative return of $\approx$ 29\% and a maximum drawdown of $\approx$ 1.8\%. Despite a smaller nominal gain relative to the S\&P 500 ($\approx$ 59\%), the strategy’s annualized Sharpe ratio remains strong at $\approx$ 2.55, roughly twice that of the benchmark ($\approx$ 1.28). The 90-day rolling Sharpe ratio continues to oscillate around elevated levels, showing that the system preserves its ability to extract cross-sectional opportunities across changing market conditions. The overall correlation to the S\&P 500 stays mildly negative ($\approx$ –6\%), indicating the portfolio continues to act as an uncorrelated alpha sleeve rather than a disguised beta exposure. The results are based on prices adjusted for corporate actions, ensuring consistency across time and preventing artificial jumps in returns due to splits, dividends, or mergers\footnote{Using adjusted prices rather than raw prices does not materially alter the outcomes; it slightly enhances reported performance by more accurately reflecting corporate actions, e.g. dividends and stock splits.}.

\begin{figure}
    \centering
    \includegraphics[trim=5mm 15mm 5mm 35mm,clip,width=\textwidth]{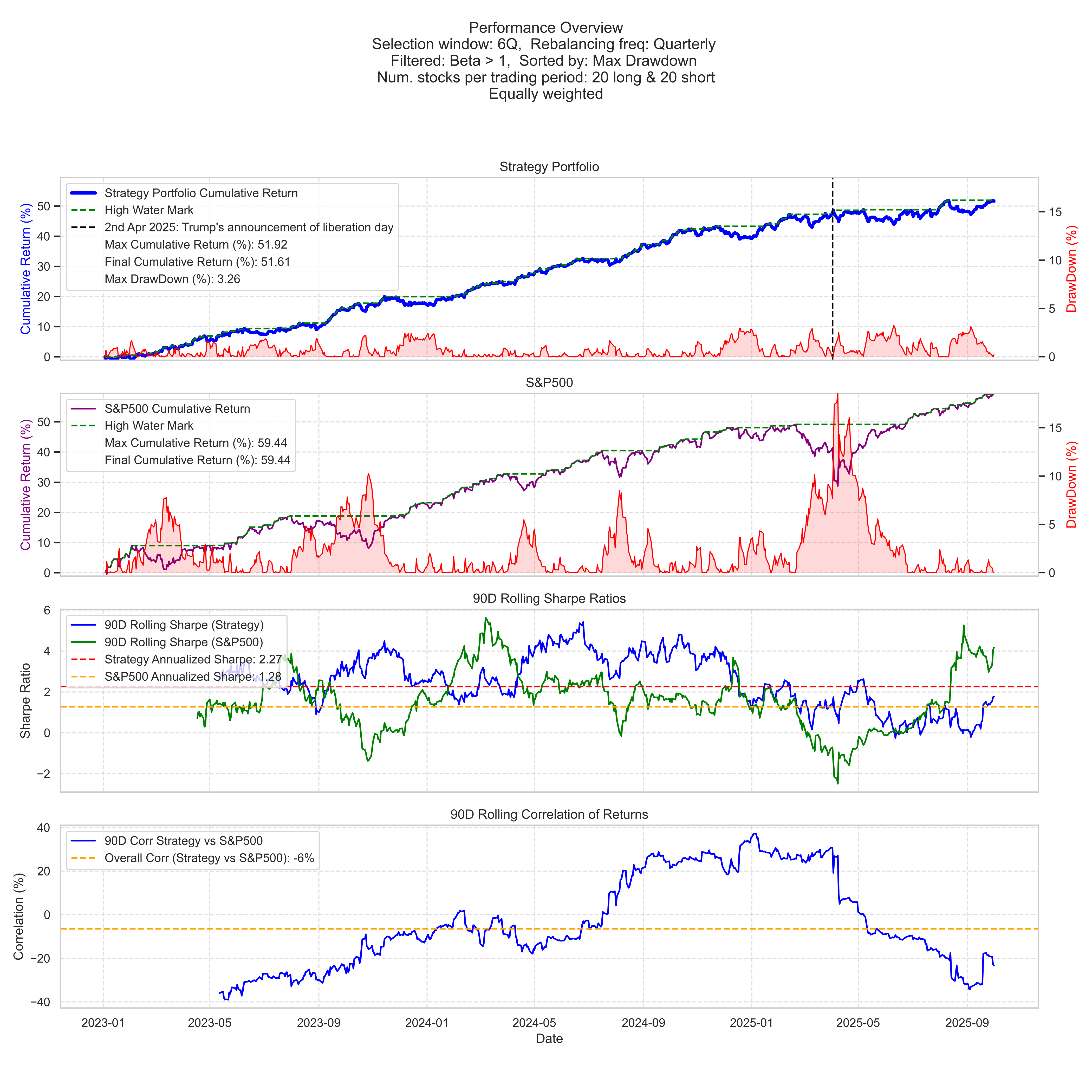}
    \caption{Out-of-sample performance of the MDD-based, equally-weighted (MDDEW) portfolio with \textbf{2x leverage and 4\% cost of capital} extended up to \textbf{Q3 2025}, benchmarked against a passive S\&P 500 buy-and-hold.}
    \label{fig:Q3_lev}
\end{figure}

For the leveraged configuration (2× exposure, 4\% cost of capital) shown in Figure \ref{fig:Q3_lev}, the cumulative return rises to $\approx$ 52\%, with a maximum drawdown of $\approx$ 3.3\% and an annualized Sharpe ratio of $\approx$ 2.27. These results reinforce the scalability of the framework: while leverage amplifies both returns and volatility, the strategy’s risk-adjusted efficiency remains well above that of the market. The extended sample through late 2025 therefore demonstrates that the signal-driven portfolio continues to deliver a low-beta, high-Sharpe return stream under diverse regimes, validating the model’s robustness and absence of performance decay over time.

\section{Conclusion}

In this paper, we presented an end-to-end evaluation of an AI-driven trading framework developed at \emph{Increase Alpha}. The model utilizes deep learning architectures—primarily feed-forward and recurrent networks—combined with expert-selected financial features to generate directional trading signals for 814 U.S. equities. The system’s architecture emphasizes operational efficiency, avoiding large-scale transformer models in favor of a leaner, interpretable design that delivers consistent performance with minimal computational overhead.

We demonstrated the predictive strength and economic viability of the model across multiple axes. Our accuracy analysis revealed that both long and short signals significantly outperform random baselines, as evidenced by statistical tests including p-values and confidence intervals. Furthermore, the signal's effectiveness persisted across various holding periods and market regimes, including the volatile period following January 2025. 

To translate these predictions into actionable trades, we conducted an extensive grid search across profit-taking, stop-loss, and holding-period parameters using Azure’s scalable cloud infrastructure. The resulting configuration enabled a robust simulation framework that quantified the strategy’s profitability and risk metrics. Notably, the framework achieved superior cumulative returns and Sharpe ratios compared to both buy-and-hold strategies and macro benchmarks like the S\&P 500, while maintaining significantly lower drawdowns.

Stress testing confirmed that the model remains stable and adapts under adverse market conditions. By modulating activity in low-confidence regimes, the framework avoids overtrading and limits exposure to high-risk periods without explicit correlation inputs. This emergent behavior highlights the system’s ability to internalize uncertainty and act accordingly.

Ultimately, our findings suggest that local inefficiencies in financial markets can be systematically harvested using targeted deep learning tools grounded in domain expertise. The \emph{Increase Alpha} platform exemplifies how rigorous engineering, paired with scalable infrastructure and disciplined evaluation, can produce a reliable and uncorrelated source of alpha. We believe this work opens the door to further applications of interpretable and computationally efficient AI in modern asset management.

%
%
\bibliographystyle{splncs04}
\bibliography{mybibliography}

\end{document}